\documentclass[twocolumn,pre,twoside,floatfix]{revtex4-1}
\usepackage{graphics,psfrag}
\usepackage{graphicx,psfrag}
\usepackage{amsmath}
\usepackage{amssymb}
\usepackage{textcomp}
\usepackage{multirow}
\usepackage{enumerate}
\usepackage{color}
\newcommand{\be}{\begin{equation}}
\newcommand{\ee}{\end{equation}}

\newcommand{\dps}{\displaystyle}

\begin{document}


\title{Order of wetting transitions in electrolyte solutions}

\author{Ingrid Ibagon}
\email{ingrid@is.mpg.de}
\author{Markus Bier}
\email{bier@is.mpg.de}
\author{S.\ Dietrich}
\affiliation
{
   Max-Planck-Institut f\"ur Intelligente Systeme, 
   Heisenbergstr.\ 3,
   70569 Stuttgart,
   Germany  and
   IV. Institut f\"ur Theoretische Physik,
   Universit\"at Stuttgart,
   Pfaffenwaldring 57,
   70569 Stuttgart,
   Germany
}

\date{\today}

\begin{abstract}

For wetting films in dilute electrolyte solutions close to charged walls we present analytic
expressions for their effective interface potentials. The analysis 
of these expressions renders the conditions under which corresponding wetting transitions 
can be first- or second-order. Within mean field theory we consider two models, one with short- and
one 
with long-ranged solvent-solvent and solvent-wall interactions. The analytic results 
reveal in a transparent way that wetting transitions in electrolyte solutions, which occur far away
from their critical point (i.e., the bulk correlation length is less than half of the Debye length)
are always first-order if the solvent-solvent and solvent-wall interactions are
short-ranged.
In contrast, wetting transitions close to the bulk critical point of the solvent (i.e., the
bulk correlation length is larger than the Debye length) exhibit the same wetting behavior as the
pure,
i.e., salt-free, solvent. If the salt-free solvent is governed by long-ranged solvent-solvent as
well as long-ranged solvent-wall interactions and exhibits critical wetting, adding salt can cause
the occurrence of an ion-induced first-order thin-thick transition which precedes the subsequent
continuous wetting as for the salt-free solvent.
\end{abstract}

\maketitle


\section{Introduction}

Recent theoretical studies of wetting phenomena in electrolyte solutions near charged walls have
focused on analyzing the influence of salt and surface charge density on the wetting behavior of
solvents \cite{Denesyuk2003,Denesyuk2004, Oleksy2009, Oleksy2010, Ibagon}. The corresponding models
share
certain common features such as the short range of the underlying non-electrostatic interaction
potentials and the mean-field character of the approaches. The model studied in Refs.
\cite{Denesyuk2003,Denesyuk2004} combines Cahn's phenomenological theory for the solvent with
the Poisson-Boltzmann theory for the ions. Within this model, ions and solvent molecules are
completely
decoupled. On the other hand, in Ref. \cite{Oleksy2009} the solvent and the ions are modeled as
hard spheres with Yukawa attraction between solvent-solvent, solvent-ion, and ion-ion pairs as well
as
Coulomb
interactions between ions. The model was studied by using classic density functional theory (DFT)
\cite{Evans}. Subsequently, the model used in Ref. \cite{Oleksy2010} includes the polar nature of
the
solvent explicitly, representing its molecules by dipolar hard spheres. In Ref. \cite{Ibagon} a
lattice
model for an electrolyte with nearest-neighbor attraction between all pairs of particles and Coulomb
interactions between ions is studied using classic DFT. Although the details of the models
used in
all these studies differ significantly, all of them agree concerning the trend that electrostatic
forces favor first-order wetting transitions. Therefore, the natural question arises whether this
observation is accidental or whether there is a deeper reason for it.

Most of the aforementioned studies are based on numerical calculations \cite{Oleksy2009,
Oleksy2010,
Ibagon} and only in Refs. \cite{Denesyuk2003,Denesyuk2004} analytic expressions for the
so-called effective
interface potential \cite{Schick,Dietrich1988}, which provides all relevant informations about
wetting transitions, have
been derived and analyzed systematically. However, this analysis is involved because it is based
on solutions of non-linear differential equations which show a complex
dependence on the relevant parameters.
 
Here, in order to infer how the order of the wetting transition is affected by the
presence of particles with electrostatic interactions, we resort to a suitable model for an
electrolyte solution near a charged wall
which has been introduced and studied in Ref. \cite{Bier2012}. Within this approach we derive an
approximate expression for the
effective interface potential, the analysis of which provides a transparent understanding of the
wetting
behavior of electrolyte solutions. In Sec. \ref{sr} we study the case of short-ranged
solvent-solvent
and short-ranged solvent-wall interactions. The case of long-ranged solvent-solvent
and long-ranged solvent-wall interactions, which has not been considered before, is discussed in
Sec. \ref{lr}. We
summarize our main results in Sec. \ref{S}.

\section{Model with short-ranged interactions}\label{sr}
We consider a model \cite{Bier2012} for an electrolyte solution in three spatial dimensions
consisting
of
solvent molecules, anions (-), and cations (+) close to a charged planar wall. Solvent particles are
assumed to have a
non-vanishing volume $a^3$ whereas the ions are considered to be point-like particles. The wall
under consideration is the $\tilde x\!-\!\tilde y$ plane at $\tilde z=0$, i.e.,
$\mathbf{\tilde r}=(\mathbf {\tilde r_{||}}\!=\!(\tilde x,\tilde y),\tilde z\!=\!0)$ which can carry
a
surface charge
density
$\tilde\sigma=\sigma e a^{-2}$, where $e>0$ is the elementary charge. We start from the following
variational grand canonical functional, which is a modification of the one
introduced in Ref. \cite{Bier2012}:
\be\label{functional}
\begin{aligned}
\beta\Omega_0[\phi(\mathbf r),\rho_\pm(\mathbf r)]&=\int d^3r\biggl\{\phi(\mathbf
r)(\ln(\phi(\mathbf
r))-\beta\mu_{\phi})\\
&+(1-\phi(\mathbf r))\ln(1-\phi(\mathbf r))\\
&+\chi(T)\phi(\mathbf r)(1\!-\!\phi(\mathbf
r))\!+\!\frac{\chi(T)}{6}(\nabla\phi(\mathbf r))^2\!\biggr\}\\
&-\beta h_1\int d^2r_{||}\phi(\mathbf {r_{||}},z=0)\\
&+\beta\frac{g}{2}\int d^2r_{||}\phi(\mathbf
r_{||},z=0)^2\\
&+\int d^3r\biggl\{\sum_{i=\pm}\rho_i(\mathbf r)\left(\ln\rho_i(\mathbf r)\right.\\
&\left.-1-\beta\mu_i+V_i(\phi(\mathbf
r))\right)\\
&+\frac{2\pi
l_B}{\varepsilon(\phi(\mathbf r))}\left( \mathbf D(\mathbf r,[\rho_\pm])\right)^2\biggr\},
\end{aligned}
\ee
where $\beta=(k_BT)^{-1}$ is the inverse thermal energy, $\mu_{\phi}$ is the chemical potential of
the solvent, $\mu_\pm$ are the chemical potentials of the $\pm$-ions,
$\tilde l_B=l_Ba=e^2\beta/(4\pi\varepsilon_0)$ is the Bjerrum length in vacuum, and $\mathbf
r=\mathbf{\tilde r}/a$ are dimensionless positions. The actual number density of the solvent is
given by $\tilde\phi(\mathbf r)=\phi(\mathbf r)a^{-3}$ with $\phi(\mathbf r)\in[0,1]$, 
whereas the number densities of anions and
cations are given by $\tilde\rho_\pm(\mathbf r)=\rho_\pm(\mathbf r)a^{-3}$. 
 In the following the fluid solvent at position $\mathbf{r}$ with $\phi(\mathbf{r})<1/2$ is
referred to as a ``gas'', whereas for $\phi(\mathbf{r})>1/2$ it is called a ``liquid''.
The first and the last integral are taken over
the half-space $\mathbf r=(x,y, z\!\ge\!0)$ whereas the second and the third integral run over the
surface $z=0$; $\rho_\pm\left(\ln\rho_\pm-1-\beta\mu_\pm\right)$
is the bulk grand potential density of the $\pm$-ions in the low number density limit. The
Flory-Huggins parameter $\chi(T)>0$ describes
the effective interaction between solvent particles \cite{Rubinstein}. The excess free
energy
of the solvent $\beta F_{ex}^{sol}[\phi(\mathbf r)]=\int d^3r\big[\chi(T)\phi(\mathbf
r)(1\!-\!\phi(\mathbf
r))\!+\!\frac{\chi(T)}{6}(\nabla\phi(\mathbf r))^2\big]$ is taken into account using the
square-gradient
approximation. The ratio $1/6$ of the coefficients in the two terms of
$\beta F_{ex}^{sol}[\phi(\mathbf
r)]$ follows from considering nearest neighbors only \cite{Cahn}. Within this model the
interaction of the
solvent with the wall is captured by the parameters $h_1$ and $g$. This implicitly assumes that
the fluid-wall interactions are sufficiently
short ranged so that their contributions to $\Omega_0$ depend only on the solvent
density $\phi(\mathbf r_{||},z\!=\!0)$ in the vicinity of the wall. This parametrization has
been used by Nakanishi and Fisher \cite{Nakanishi} in order to
analyze the global surface phase diagram of the Landau-Ginzburg theory for wetting. 
 $V_\pm(\phi)$ is  the solvation free energy per $k_BT$ of a $\pm$-ion 
 in the solvent  of number density $\phi$.
Whereas more realistic expressions of $V_\pm(\phi)$ are discussed in the literature
\cite{Bier2012},
we use here a simple piece-wise constant expression $V_\pm(\phi<1/2)=V_g$ and
$V_\pm(\phi>1/2)=V_l$ with
$V_g-V_l\gg1$. 
This choice guarantees a vanishingly small ionic strength in the gas ($\phi<1/2$) as compared
to the
ionic strength in the liquid ($\phi>1/2$).
Without restriction of generality we choose $V_l:=0$, which can be achieved by a redefinition of the ionic
chemical potentials ($\beta\mu_\pm-V_l\mapsto\beta\hat{\mu}_\pm$; in the following we drop
the hat $\hat{}$).
The discontinuity of $V_\pm(\phi)$ at $\phi=1/2$ is expected to not affect the results
significantly because only thermodynamic states of liquid-gas coexistence well below the critical
point are considered, for which $\phi=1/2$ is deep inside the unstable region of the bulk phase
diagram. Note that here no unequal partitioning of ions in a non-uniform solvent occurs due to
$V_+(\phi)-V_-(\phi)=0$,
i.e., due to a vanishing difference of solubility contrasts of anions and cations between the
two phases in the sense of Ref.~\cite{Bier2012}.
Moreover, no specific adsorption of ions at interfaces is considered here, i.e., there are no
surface fields acting on $\rho_\pm$.
$\tilde {\mathbf D}=\mathbf{D}ea^{-2}$ is the electric displacement generated by the ions and by the
surface charge density as related according to Gau{\ss}'s law $\nabla \cdot \mathbf{D}(\mathbf
r,[\rho_\pm])=\rho_+(\mathbf r)-\rho_-(\mathbf r)+\sigma\delta(z)$. (Note that Gau{\ss}'s law
is an ingredient of the theory in addition to Eq. (\ref{functional}).) Within the present model,
ions
interact among each other and with the wall only electrostatically (besides the hard core
repulsion of the wall which prevents the ions to penetrate the wall). Here, this is expressed in
terms of the energy density of the electric field \cite{Ibagon} where
$\varepsilon(\phi)$ is the local permittivity of
the solvent of density $\phi$ divided by the vacuum permittivity $\varepsilon_0$.
Various empirical expressions for $\varepsilon(\phi)$ are in use \cite{Boettcher1973}.
However, for the sake of simplicity here we adopt a simple piece-wise constant expression 
$\varepsilon(\phi<1/2)=1$ and $\varepsilon(\phi>1/2)=\varepsilon_l$ with the relative
permittivity $\varepsilon_l$
of the liquid solvent.
For the same reasons as for the case of the piece-wise constant expressions $V_\pm(\phi)$ (see above),
the discontinuity of $\varepsilon(\phi)$ at $\phi=1/2$  is expected to be irrelevant for the
present purposes.

The \textit{b}ulk grand canonical potential density per $k_BT$ following from Eq.~(\ref{functional})
is
given by
\be\label{omegabulk}
\begin{aligned}
\beta\Omega_b(\phi,\rho)&=f_{sol}(\phi)+f_{ion}^{(+)}(\rho)+f_{ion}^{(-)}(\rho)\\
&+\rho\left(V_+(\phi)+V_-(\phi)\right)
\end{aligned}
\ee
with $\rho_+=\rho_-:=\rho$ due to local charge neutrality in the bulk, and with the abbreviations
$f_{sol}(\phi)\!:=\!\phi(\ln(\phi)\!-\!\beta\mu_{\phi})+(1\!-\!\phi)\ln(1\!-\!\phi)\!+\!\chi(T)\phi(1\!-\!\phi)$
and $f^{(\pm)}_{ion}(\rho_\pm)\!:=\!\rho_\pm\left(\ln\rho_\pm\!-\!1\!-\!\beta\mu_\pm\right)$. As a
consequence of local charge neutrality $\Omega_b$ depends on $\mu_+$ and $\mu_-$ only via the
 combination $\mu_++\mu_-\equiv\mu_I$. Accordingly, the ionic chemical potentials $\mu_\pm$
are of no individual importance but only their sum is of physical relevance.
In the bulk $D=0$ due to local charge neutrality so that the last term in Eq.~(\ref{functional}) does not 
contribute to Eq.~(\ref{omegabulk}).
Equilibrium bulk states $(\phi, \rho)$ minimize $\beta\Omega_b(\phi,\rho;\mu_{\phi}, \mu_I,
T)$, i.e., they fulfill the
Euler-Lagrange equations
\be
\frac{\partial\Omega_b}{\partial\phi}=0 
\label{eulerphi}
\ee
and
\be
\frac{\partial\Omega_b}{\partial\rho}=0.
\label{eulerI}
\ee
Equations (\ref{eulerphi}) and (\ref{eulerI}) render two solutions, i.e., minima:
$\left[\phi_l(\mu_\phi,\mu_I, T),\rho_l(\mu_\phi,\mu_I, T)\equiv I\right]$ and
$\left[\phi_g(\mu_\phi,\mu_I, T),\rho_g(\mu_\phi,\mu_I, T)\right]$. Coexistence between
these two minima occurs if upon inserting these two solutions into $\Omega_b$ the minima are equally
deep:
\be\label{coex}
\begin{split}
\Omega_b\left(\phi=\phi_l(\mu_\phi,\mu_I, T),\rho=\rho_l(\mu_\phi,\mu_I,
T);\mu_\phi,\mu_I, T\right)\\
=\Omega_b\left(\phi=\phi_g(\mu_\phi,\mu_I,
T),\rho=\rho_g(\mu_\phi,\mu_I,
I);\mu_\phi,\mu_I, T\right). 
\end{split}
\ee 
This renders a relation $\mu_\phi=\mu_\phi^{co}\left(\mu_I,T\right)$ which describes a
two-dimensional manifold in the three-dimensional parameter space $\left(\mu_\phi,\mu_I, T\right)$
where gas-liquid coexistence occurs. Inserting this relations into the solutions renders
$\left[\phi^{co}_l(\mu_I, T),\rho^{co}_l(\mu_I,T)\right]$ and $\left[\phi^{co}_g(\mu_I,
T),\rho^{co}_g(\mu_I,T)\right]$ with
$\phi^{co}_{l,g}\left(\mu_I,T\right)=\phi_{l,g}\left(\mu_\phi=\mu_\phi^{co}\left(\mu_I,T\right)
, \mu_I , T\right)$ and
$\rho^{co}_{l,g}\left(\mu_I,T\right)=\rho_{l,g}\left(\mu_\phi=\mu_\phi^{co}\left(\mu_I,T\right)
,\mu_I,T\right)$
where $\rho^{co}_{l}=I$. In order
to avoid a clumsy notation, in the following we drop the superscript \textit{co} in $\phi_l^{co}$,
$\phi_g^{co}$, and $\rho_g^{co}$ so that, if not stated otherwise, $\phi_l$, $\phi_g$, $I$, and
$\rho_g$ correspond to the coexisting densities.

Equation (\ref{eulerI}) can be used to express the bulk ionic strength as
\be\label{I}
\rho=\exp\left(\frac12\left(\beta\mu_I-V_+(\phi)-V_-(\phi)\right)\right).
\ee
Due to our choice $V_\pm(\phi<1/2)=V_g$ and $V_\pm(\phi>1/2)=V_l=0$ we obtain for the ionic
strength in the
gas ($\phi=\phi_g<1/2$) $\rho=\rho_g=\exp(\beta\mu_I/2-V_g)$ and in the liquid
($\phi=\phi_l>1/2$) 
$\rho=\rho_l=\exp(\beta\mu_I/2)$ (see Eq. (\ref{I})).
In the following it is assumed that $V_g\gg1$ such that we can set $\rho_g=0$, and
$I=\rho_l=\exp(\beta\mu_I/2)$. Accordingly, by using $\mu_I=2k_BT\ln I$ the densities discussed
above can be expressed as functions of $I$ and $T$. Note that Eq. (\ref{I}) is independent of the
Flory-Huggins
parameter $\chi(T)$. For our choice of the ion potential $V_\pm(\phi)$ (such that
$V'_\pm(\phi\not=1/2)=V''_\pm(\phi\not=1/2)=0$)
the binodal $T_{bi}(\phi)$ is determined via the implicit relation 
\be\label{binodal}
\chi(T)=\frac{\ln(\phi)-\ln(1-\phi)}{2\phi-1},
\ee
where the temperature dependence of $\chi(T)$ is often taken to be 
$\chi(T)\cong\chi_S+\chi_H/T$,
 where $\chi_S$ and $\chi_H/T$ are referred to as the entropic and the enthalpic part of
$\chi(T)$,
respectively \cite{Rubinstein}. From Eq.~(\ref{binodal}) one infers the critical point to be located
at $(\phi_c=1/2,\ \chi(T_c)=2)$.
Note that within our approximation the binodal (and hence the critical point) is independent of the
ionic strength $I$.

In the presence of walls, $\phi$ and $\rho_\pm$ vary spatially in normal direction $z$.
Their equilibrium profiles minimize the full functional $\Omega_0[\phi(\mathbf
r),\rho_\pm(\mathbf r)]$
in Eq.~(\ref{functional}) and thus render the equilibrium state. This
procedure can be performed
numerically.
However, for the present purpose, we seek analytic expressions. In order to achieve
this goal we perform a Taylor expansion of the local part in Eq.~(\ref{functional})
around  the \textit{s}harp-\textit{k}ink reference density profiles \cite{Dietrich1988}
\be\label{sharpkinkphi}
\bar\phi(z)=\phi_{sk}(z)=\begin{cases}
  \phi_l, & 0\le z\le \ell\\
\phi_g, &  z>\ell    
        \end{cases}
\ee
and  
\be
\bar\rho_{\pm}(z)=\rho_{sk,\pm}(z)=\begin{cases}
  I, & 0\le z\le \ell \\
0, & z>\ell     
        \end{cases}
\label{sharpkinkI}
\ee 
where $\ell$ is the position of the discontinuity of the sharp-kink profile $\phi_{sk}(z)$,
and $\phi_l$ and $\phi_g$ are,
respectively, the
equilibrium bulk densities of the solvent in the liquid and gas phase for a bulk ionic
strength $I$ in the liquid phase. 
This Taylor expansion
renders an approximate variational functional $\hat \Omega_0$ which up to quadratic order is
given by

\be\label{functional_approx}
\begin{aligned}
\frac{\beta\hat\Omega_0[\phi(z),\rho_\pm(z)]}{A}&=\ell f_{sol}
(\phi_l)+(L-\ell)f_{sol}(\phi_g)\\
&+\int_0^L
dz\Bigg\{ f'_{sol}\left(\bar\phi(z)\right)(\phi(z)-\bar\phi(z))\\
&+\frac12 f''_{sol}\left(\bar\phi(z)\right)(\phi(z)-\bar\phi(z))^2\\
&+\frac{\chi(T)}{6}\left(\frac{d\phi(z)}{dz}\right)^2\Bigg\}\\
&-\beta h_1\phi(0) + \beta\frac{g}{2}(\phi(0))^2\\
&+\int^\ell_0\!\!dz\Bigg\{\sum_{i=\pm}
\Big[f^{(i)}_{ion}(I)\\
&+\!{f^{(i)}_{ion}}'\!(I)(\rho_i(z)\!-\!I)\\
&+\frac12{f^{(i)}_{ion}}''(I)(\rho_i(z)-I)^2\Big]\\
& +\frac{2\pi l_B}{\varepsilon_l}\left(D(z,[\rho_\pm])\right)^2\Bigg\},
\end{aligned}
\ee
where $\tilde A=Aa^2$ is the wall area and $\tilde V= ALa^3$ is the volume of the system. In
order to
obtain Eq. (\ref{functional_approx}) it has been used that $\rho_\pm(z>\ell)=0$, $V_\pm(\phi(z\leq
\ell))=V_l=0$ because $\phi(z\leq \ell)>1/2$, and
$\varepsilon(\phi\leq \ell)=\varepsilon_l$. Therefore  Eq.~(\ref{functional_approx})
does not apply very close to the critical point where the actual spatial variation of
$V_{\pm}(\phi(z))$ and $\varepsilon(\phi(z))$ matters.
Moreover, $D(z>\ell)=0$  because the gas phase contains no ions and $D(z\to\infty)\to0$ due
to the constraint of global charge neutrality.

The Euler-Lagrange equation for $\phi(z)$, which follows from Eq.~(\ref{functional_approx})
for fixed $\ell$, is given by

\be\label{eqphiz}
\begin{aligned}
\frac{\chi(T)}{3}\frac{d^2\phi(z)}{dz^2}
=f'_{sol}\left(\bar\phi(z)\right) + f''_ {sol}\left(\bar\phi(z)\right)(\phi(z)-\bar\phi(z)) \\
\end{aligned}
\ee
with the boundary conditions
\be\label{bc1}
\frac{\chi(T)}{3}\left.\frac{d\phi(z)}{dz}\right|_{z=0}=-\beta h_1+\beta g\phi(0)\ \ \text{and}\ \
\ \left.\frac{d\phi(z)}{dz}\right|_{z=L}=0.
\ee
Similarly,  the Euler-Lagrange equations for $\rho_\pm(z), 0\leq z\leq\ell,$ read (using
Eq.~(\ref{ED}) and 
$\frac{d}{dz}\frac{\delta D(z)}{\delta \rho_i(z')}=\frac{\delta }{\delta
\rho_i(z')}\frac{dD(z)}{dz}=q_i\delta(z-z')$ due to Gau\ss's law)
\be\label{eqions}
{f^{(i)}_{ion}}'(I) + {f^{(i)}_{ion}}''(I)(\rho_i(z)-I)=-q_i\Psi(z),
\ee
where $eq_\pm$ is the ion charge with $q_{\pm}=\pm1$ and
$\tilde\Psi(z)=\Psi(z)/(\beta e)$ is
the electrostatic potential such that
\be\label{ED}
D(z)=-\frac{\varepsilon_l}{4\pi l_B}\Psi'(z)
\quad\text{for $0\leq z \leq\ell$.}
\ee
The variation leading to Eq. (\ref{eqions}) generates also boundary terms $\Psi(z)\delta
D(z,[\rho_\pm])/\delta\rho_i(z')$ at $z=0$ and $z=\ell$ which, however, vanish because of the
boundary conditions
$D(z=0)=\sigma$  and $D(z=\ell)=0$. The latter holds due to $D\equiv0$ in the gas and the continuity
of
$D(z)$ at $z=\ell$ in the absence of a surface charge at $z=\ell$. Due to Eq. (\ref{eulerI}), in
Eq. (\ref{eqions}) one has ${f_{ion}^{(i)}}'=0$. For the particular choice of $\bar\phi(z)$  in Eq.
(\ref{sharpkinkphi}) one has $f'_{sol}(\bar\phi(z<\ell))=f'_{sol}(\phi_{l})=0$ and
$f'_{sol}(\bar\phi(z>\ell))=f'_{sol}(\phi_{g})=0$ due to Eq. (\ref{eulerphi}) and the
Euler-Lagrange
equation (\ref{eqphiz}) can be written as
\be\label{eqphi}
\frac{d^2}{dz^2}\Delta\phi(z)=\begin{cases}\frac{1}{\xi_l^2}\Delta\phi(z),&  0\le z< \ell\\
\frac{1}{\xi_g^2}\Delta\phi(z),&  z> \ell
\end{cases}
\ee
 with
\be\label{xi}
 \frac{1}{\xi_{g,l}^2}=\frac{3}{\chi(T)}\left(\frac{1}{\phi_{g,l}}+\frac{1}{1-\phi_{g,l}}
-2\chi(T)\right),
\ee
where $\xi_{g,l}$ can be identified with the bulk correlation length of the solvent in the gas
and in the liquid phase, respectively (see Appendix \ref{axi}), and where
$\Delta\phi(z)=\phi(z)-\bar\phi(z)=\phi(z)-\phi_l$
for
$0\le
z<\ell$ and $\Delta\phi(z)=\phi(z)-\phi_g$ for
$z>\ell$. In addition to the boundary conditions in Eq. (\ref{bc1}), Eq.~(\ref{eqphiz})
requires
the density profile $\phi(z)$ and its first derivative $\frac{d\phi(z)}{dz}$ to be continuous at $z=\ell$, i.e.,  
\be\label{BCphi}
\begin{aligned}
\phi_l+\Delta\phi(\ell^-)&=\phi_g+\Delta\phi(\ell^+)\\
\left.\frac{d}{dz}\Delta\phi(z)\right|_{z=\ell^-}&=\left.\frac{d}{dz}\Delta\phi(z)\right|_{z=\ell^+}
\end{aligned}
\ee
 because the right-hand side of Eq. (\ref{eqphiz}) is discontinuous; otherwise the left-hand
side of Eq.~(\ref{eqphiz}) would be more singular at $z=\ell$
than the right-hand side.
Similarly, for $\Delta\rho_\pm(z)$ one obtains (see Eqs.~(\ref{eulerI}) and (\ref{eqions}))
\be\label{deltaions}
\Delta\rho_\pm(z)=-q_\pm\Psi(z)I,
\ee
where $\Delta\rho_\pm(z)=\rho_\pm(z)-\bar\rho_\pm(z)=\rho_\pm(z)-I$ for $0\le z < \ell$ and
zero otherwise (see Eq.
(\ref{sharpkinkI}) and $\rho_\pm(z>\ell)=0$). The dependences of Eqs. (\ref{eqphiz}) and
(\ref{deltaions}) on $\mu_\phi$ and
$\mu_I$ enter via the equilibrium values
of $I$, $\phi_g$, and $\phi_l$ (Eqs. (\ref{omegabulk})-(\ref{coex})).

The Poisson equation, which relates the dimensionless electrostatic potential
$\Psi=\beta e\tilde\Psi$ to
the dimensionless number densities $\rho_\pm$ of the ions, can be written as (see Eqs.
(\ref{ED}) and
(\ref{deltaions}) and Gau\ss's law)
\be\label{LPB}
\begin{aligned}
 \Psi''(z)=-\frac{4\pi\lambda_B}{\varepsilon_l}\frac{dD}{dz}&=-\frac{4\pi
l_B}{\varepsilon_l}\sum_{i=\pm}q_i\rho_i(z)\\
&=-\frac{4\pi l_B}{\varepsilon_l}\sum_{i=\pm}q_i(\Delta\rho_i+I)\\
&=-\frac{4\pi l_B}{\varepsilon_l}\sum_{i=\pm}q_iI(1-q_i\Psi(z))\\
&=\frac{8\pi Il_B}{\varepsilon_l}\Psi(z)\\
&=\kappa^2\Psi(z)
\end{aligned}
\ee
which is the linearized Poisson-Boltzmann equation with 
\be\label{kappa}
\kappa=\sqrt{8\pi Il_B/\varepsilon_l}
\ee
as the inverse Debye length. This equation must be solved subject to the boundary
conditions
\be\label{BCLPB}
\Psi'(z=0)=-\frac{4\pi l_B\sigma}{\varepsilon_l}\ \ \text{and}\ \ \Psi'(z=\ell)=0
\ee
which corresponds to $D(z=0)=\sigma$ and $D(z=\ell)=0$; the latter follows from the overall
charge neutrality (see Eqs. (13) and (14) in Ref. \cite{Ibagon})
and the assumption that there are no ions in the vapor phase.
Equations (\ref{eqphi}) and (\ref{LPB}) can be solved
analytically, yielding the constrained equilibrium profiles
\be\label{solutionphi}
\Delta\phi^{(\ell)}(z)=\begin{cases}
               
\hat A_l\exp(z/\xi_l)+\hat B_l\exp(-z/\xi_l), & 0\le z\le \ell\\
\hat B_g\exp(-z/\xi_g), &  z>\ell
              \end{cases}
\ee
where (see Eqs.~(\ref{BCphi}) and (\ref{bc1}))
\be\label{consts}
\begin{aligned}
\hat A_l&=\frac{1}{N}\left[\left(\beta g+\frac{\chi(T)}{3\xi_l}
\right)(\phi_g-\phi_l)\right.\\
&\left.+\beta(h_1-g\phi_l)\left(\frac {
\xi_g } { \xi_l } -1\right)\exp(- \ell/\xi_l) )\right ],\\
\hat B_l&=\frac{1}{N}\left[\left(\frac{\chi(T)}{3\xi_l}
-\beta g\right)(\phi_g-\phi_l)\right.\\
&\left.+\beta(h_1-g\phi_l)\left(1+\frac { \xi_g } { \xi_l
}\right)\exp(\ell/\xi_l) )\right ],\\
\hat B_g&=\frac{\xi_g}{\xi_l}\left(B_l\exp\left(\frac{\ell}{\xi_g}-\frac{\ell}{\xi_l}\right
)-\hat A_l\exp\left(\frac{\ell}{\xi_g}+\frac{\ell}{\xi_l}\right)\right),\\
\hat N&=\left(\beta g+\frac{\chi(T)}{3\xi_l}\right)\left(\frac{\xi_g}{\xi_l}
+1\right)\exp(\ell/\xi_l)\\
&-\left(\beta g-\frac{\chi(T)}{3\xi_l}\right)\left(1-\frac{\xi_g}{\xi_l}
\right)\exp(- \ell/\xi_l),
\end{aligned}
\ee
and (see Eq. (\ref{deltaions})) $\Delta\rho_\pm^{(\ell)}(z)=-q_\pm I\Psi^{(\ell)}(z)$ with
\be\label{solutionEP}
\Psi^{(\ell)}(z)=A_I\exp(\kappa z)+B_I\exp(-\kappa z),
\ee
where (see Eq. (\ref{BCLPB}))
\be\label{const-ions}
\begin{aligned}
 A_I&=\frac{4\pi l_B\sigma}{\varepsilon_l\kappa}\frac{1}{\exp(2\kappa \ell)-1},\\
 B_I&=A_I\exp(2\kappa \ell).
\end{aligned}
\ee
 Note that because $\rho_\pm(z)\ge0$, $|\Delta\rho_\pm(z)|$ has an upper limit given by (see Eq.
(\ref{deltaions}))
\be
|\Delta\rho_\pm(z)|\le I\text{, i.e.,}\ |\Psi(z)|\le 1,
\ee
Since $\Psi(z)$ is monotonic (see Eqs. (\ref{solutionEP}) and (\ref{const-ions})) one requires
\be
\begin{aligned}
 |\Psi(0)|\le 1\\
\Rightarrow \frac{4\pi l_B|\sigma|}{\varepsilon_l\kappa}\coth(\kappa\ell)\le 1,
\end{aligned}
\ee
which implies that there is an upper limit for the absolute value $|\sigma|$ surface charge
density: 
\be\label{limitsigma}
\begin{aligned}
|\sigma|&\le\frac{\varepsilon_l\kappa}{4\pi l_Bq_\pm}\tanh(\kappa\ell)=:\sigma_{sat}(\kappa\ell).\\
\end{aligned}
\ee
In case the real surface charge $|\sigma|$ is larger than the saturation value
$\sigma_{sat}(\kappa\ell)$ the latter has to be used instead in order to ensure
$\rho_{\pm}(z)\ge0$. This is the analogue of the well-known charge renormalization in the
linearized Poisson-Boltzmann theory of semi-infinite electrolyte solutions \cite{Bocquet}.

Within the present model, at two-phase coexistence $\phi_g=1-\phi_l$ so that
$\xi_g=\xi_l=\xi$ and that Eq. (\ref{consts}) reduces to
\be\label{consts_co}
\begin{aligned}
A_{l}&=\frac{1}{N}\left(\beta g+\frac{\chi(T)}{3\xi}
\right)(\phi_g-\phi_l),\\
B_{l}&=\frac{1}{N}\biggl[\left(\frac{\chi(T)}{3\xi}
-\beta g\right)(\phi_g-\phi_l)\\
&+2\beta(h_1-g\phi_l)\exp(\ell/\xi) )\biggr],\\
B_{g}&=\left(B_{l}-A_{l}\exp(2 \ell/\xi)\right),\\
N&=2\left(\beta g+\frac{\chi(T)}{3\xi}\right)\exp(\ell/\xi).
\end{aligned}
\ee
The solvent density profiles at two-phase coexistence --- obtained by expanding $\Omega_0$ up to
quadratic order around the sharp-kink profiles (see Eqs.  (\ref{solutionphi}) and
(\ref{consts_co})) --- are similar but not identical to the ones obtained by using the so-called
double
parabola approximation (DPA) (see Appendix \ref{DPS}). A difference between the two approaches
arises
because the boundary conditions and the definition of the thickness of the liquid film differ in
both approaches. Within the present approach the thickness $\ell_{film}$ of the liquid film is
defined
as the
position $z>0$ in which the magnitude of the derivative $|\phi'(z)|$ of the solvent profile is
maximal. This definition of the film thickness $\ell_{film}$ is convenient within the present
approach because Eq. (\ref{solutionphi}) shows that it coincides with the
position $\ell$ of the discontinuity of the sharp-kink
profile $\phi_{sk}(z)$, i.e., $\ell_{film}=\ell$. Alternative definitions of the film
thickness are possible but lead to more complicated expressions of the effective interface potential
introduced below. Moreover, the present solvent profile and its derivative are continuous
everywhere (see Eq. (\ref{BCphi})), whereas within the DPA the solvent profile is
continuous everywhere but its derivative is discontinuous at the position $z=\ell_{\text{DP}}$
where $\phi_{\text{DP}}(\ell_{\text{DP}})=\frac 12(\phi_g+\phi_l)$ (see Eq. (\ref{SDP})).
This equation defines the liquid film thickness $\ell_\text{DPA}$ within the DPA. 
However, the discontinuity
$\phi'_{\text{DPA}}(\ell_{\text{DPA}}^+)-\phi'_{\text{DPA}}(\ell_{\text{DPA}}^-)
=\mathcal{O}(\exp(-\ell_\text{DP}/\xi))$ is exponentially small for large film thicknesses
($\ell_\text{DPA}\gg\xi$)
(see Eqs.~(\ref{SDP}) and (\ref{DPcoeff})).
Moreover, in Appendix~\ref{DPS} it is shown that the relative difference between the
coefficients of the profiles in 
Eq.~(\ref{solutionphi}) and those in Eq.~(\ref{SDP}) is also exponentially small for
$\ell=\ell_\text{DPA}\gg\xi$ 
(see Eq.~(\ref{relativedifference})).

At the functional minimum the integrations in Eq. (\ref{functional_approx}) can be performed
analytically with the first integrand (see  Eqs. (\ref{eqphi}) and (\ref{xi})) 
\be
\begin{aligned}
&\frac{\chi(T)}{6}\!
\left(\frac{d}{dz}\Delta\phi(z)\!\right)^2\!\!\!
+\!\frac12\left(\Delta\phi(z)\right)^2\!\left(\frac{1}{\phi_{g,l}}\!+\!\frac { 1 } { 1-\phi_{g,l}
}\!
-\!2\chi(T)\!\right)\\
&\qquad\qquad =\frac{\chi(T)}{6}\left(\frac{\left(\Delta\phi(z)\right)^2}{\xi_{g,l}^2}+\left(\frac{d
} { dz }\Delta\phi(z)\right)^2\right)\\
&\qquad\qquad=\frac{\chi(T)}{6}\left(\frac{\left(\Delta\phi(z)\right)^2}{\xi_{g,l}^2}
+\frac{d
} { dz }\left(\Delta\phi(z)\frac{d
} { dz }\Delta\phi(z)\right)\right.\\
&\qquad\qquad\left.-\Delta\phi(z)\frac{d^2
} { dz^2
}\left(\Delta\phi(z)\right)\vphantom{\frac{\left(\Delta\phi(z)\right)^2}{\xi_{g,l}^2}}\right)\\
&\qquad\qquad=\frac{\chi(T)}{6}\frac{d
} { dz }\left(\Delta\phi(z)\frac{d
} { dz }\Delta\phi(z)\right),
\end{aligned}
\ee
and with the second the integrand (see Eqs.~(\ref{ED}), (\ref{deltaions}), and (\ref{LPB}) and
${f_{ion}^{(i)}}'(I)=0$)
\begin{align}
&\frac{1}{2I}\sum_{i=\pm}\left(\Delta\rho_i(z)\right)^2+\frac{2\pi
l_B}{\varepsilon_l}\left(D(z,[\Delta\rho_\pm])\right)^2\nonumber\\
&\qquad\qquad\qquad=\frac{I}{\kappa^2}
\left(\kappa^2\left(\Psi(z)\right)^2+\left(\Psi'(z)\right)^2\right)\nonumber\\
&\qquad\qquad\qquad=\frac{I}{\kappa^2}
\left(\Psi(z)\Psi''(z)+\left(\Psi'(z)\right)^2\right)\nonumber\\
&\qquad\qquad\qquad=\frac{I}{\kappa^2}\left(\Psi(z)\Psi'(z)\right)'.
\end{align}
By exploiting the boundary conditions in Eqs. (\ref{bc1}), (\ref{BCphi}), and (\ref{BCLPB}) one
obtains for the surface contribution to the constrained grand potential
\be\label{Eqomegas}
\begin{aligned}
\beta\Omega_s(\ell)&:=\frac{\beta\hat\Omega_0[\phi^{(\ell)}(z),\rho^{(\ell)}_\pm(z)]-V\beta\Omega_b(\phi_g,\rho_g=0)}{A}\\
&=\beta\ell[\Omega_b(\phi_l,I)-\Omega_b(\phi_g,0)]\\
&\phantom{=}+\frac{\chi(T)}{6}\big( \Delta{\phi^{(\ell)}}'(\ell)(\phi_g\!-\!\phi_l)\!-\!
\Delta{\phi^{(\ell)}}'(0)\Delta\phi^{(\ell)}(0)\big)\\
&\phantom{=}-\beta h_1(\phi_l+\Delta\phi^{(\ell)}(0))+\frac{\beta g}{2}(\phi_l+\Delta\phi^{(\ell)}(0))^2\\
&\phantom{=}+\frac{1}{2}\sigma\Psi^{(\ell)}(0)
\end{aligned}
\ee
where
$\Delta{\phi^{(\ell)}}'(\ell)=\left.\left(\frac{d}{dz}\Delta\phi^{(\ell)}(z)\right)\right|_{z=\ell}
$. Finally,
inserting the solutions given by Eqs.~(\ref{solutionphi})-(\ref{const-ions}) into
Eq.~(\ref{Eqomegas}) and using 
Eqs.~(\ref{bc1}) and (\ref{BCLPB}) leads to
\be\label{omegas}
\begin{aligned}
\beta\Omega_s(\ell) &= \beta\ell[\Omega_b(\phi_l,I)-\Omega_b(\phi_g,0)]\\
&+\frac{\chi(T) }
{6}(\phi_g-\phi_l)\biggl[\frac{\hat A_l}{\xi_l}\exp(\ell/\xi_l)\\
&-\frac{\hat B_l}{\xi_l}\exp(- \ell/\xi_l)\biggr]
\\
&+\frac{\beta(g\phi_l-h_1)}{2}(\hat A_l+\hat B_l)-\beta h_1\phi_l+\frac{\beta g}{2}\phi_l^2\\
&+\frac{2\pi l_B\sigma^2}{\varepsilon_l\kappa}\coth(\kappa \ell).
\end{aligned}
\ee 
The first term is the difference between the grand canonical potentials per volume of the
(potentially metastable) liquid and the
gas bulk phase, respectively, multiplied by the film thickness $\ell$. This 
contribution linear in $\ell$ vanishes at two-phase coexistence due to Eq.~(\ref{coex}).
The other terms provide the free energy associated with the emergence of the liquid-gas and
the liquid-wall interfaces as well as their effective interaction for $\ell<\infty$. These
expressions are valid also off two-phase coexistence.

\textit{At} two-phase coexistence and in the limit $\ell\gg1/\kappa$ the surface contribution (Eq.
(\ref{omegas})) can be written as
\be\label{omegas_co}
\begin{aligned}
\beta\Omega_s(\ell)&\simeq   \frac{\chi(T) }
{12}\frac{(\phi_g-\phi_l)^2}{\xi}-\frac{\beta^2(h_1-g\phi_l)^2}{2\left(\beta g+\frac{\chi(T)}{3\xi}
\right) }\\
& -\beta h_1\phi_l+\frac{\beta g}{2}\phi_l\\
&+(\phi_l-\phi_g)\frac{\chi(T)}{3\xi}\frac{\beta(h_1-g\phi_l)}{\beta g+\frac{\chi(T)
} { 3\xi}}\exp(- \ell/\xi)
\\
&-(\phi_l-\phi_g)^2\frac{\chi(T)}{12\xi}\frac{\frac{\chi(T)
} { 3\xi}-\beta g}{\beta g+\frac{\chi(T)
} { 3\xi}}\exp(-2 \ell/\xi)\\
&+\frac{2\pi l_B\sigma^2}{\varepsilon_l\kappa}(1+2\exp(-2\kappa \ell)),
\end{aligned}
\ee
where only the last term in Eq. (\ref{omegas}) has been expanded for $\kappa\ell\gg1$. Note
that within  the present theory the ions enter into $\Omega_s(\ell)$ only via the last term.
Therefore, if
the surface charge $\sigma$ is zero, the ions do not modify the wetting behavior of the solvent.
This
is due to the fact that $\sigma=0$  implies that there are no surface fields acting on
$\rho_\pm$ . The
first
term in
Eq. (\ref{omegas_co}) is the liquid-gas surface tension $\gamma_{l,g}=\frac{\chi(T) }
{12}\frac{(\phi_g-\phi_l)^2}{\xi}$, within the
present approach. As expected, within mean field theory (MFT) $\phi_l-\phi_g$ vanishes
$\propto|\chi-\chi_c|^\beta$ with $\beta=1/2$ upon approaching the critical point
($\phi_c=\frac12,\chi_c=2$) (see Eq. (\ref{binodal})) and from Eq. (\ref{xi}) one
has $\xi\propto |\chi-\chi_c|^{-\nu}$ with $\nu=1/2$, so that 
$\gamma_{l,g}\propto|\chi-\chi_c|^{\mu}$ with $2\beta+\nu=\mu=3/2$; in general
$\mu=(d-1)\nu$ where $d$ is the spatial dimension with $d=4$ corresponding to MFT \cite{Schick}.
The wall-liquid surface tension is
$\gamma_{w,l}=-\frac{\beta^2(h_1-g\phi_l)^2}{2\left(\beta
g+\frac{\chi(T)}{3\xi}\right) }-\beta h_1\phi_l+\frac{\beta
g}{2}\phi_l+\frac{2\pi l_B\sigma^2}{\varepsilon_l\kappa}$. These
two contributions are independent of the film
thickness $\ell$. The remaining terms carry the dependence on $\ell$,
generated by the effective interaction between the emerging liquid-vapor and wall-liquid
interfaces.

Accordingly, the effective interface potential
$\omega(\ell)=\Omega_s(\ell)-\Omega_s(\infty)$ at
two-phase coexistence is given by
\be\label{omegacompleto}
\begin{aligned}
\beta\omega(\ell\gg1/\kappa)& \simeq 
(\phi_l-\phi_g)\frac{\chi(T)}{3\xi}\frac{\beta(h_1-g\phi_l)}{\beta
g+\frac{\chi(T)
} { 3\xi}}\exp(- \ell/\xi)\\
&-(\phi_l-\phi_g)^2\frac{\chi(T)}{12\xi}\frac{\frac{\chi(T)
} { 3\xi}-\beta g}{\beta g+\frac{\chi(T)
} { 3\xi}}\exp(-2 \ell/\xi)\\
&+\frac{4\pi l_B\sigma^2}{\varepsilon_l\kappa}\exp(-2\kappa \ell).
\end{aligned}
\ee
This effective potential captures the dependence of the grand canonical potential on
the film thickness and determines whether or not the wall-gas interface is wetted by the liquid.
Moreover, the order of the wetting transition can be inferred from its functional form
\cite{Dietrich1988}.

The property $\xi_l=\xi_g$ at coexistence is a special feature of the present model. In general
$\xi_l\neq\xi_g$ so that in this case an expansion of the effective
interface potential $\omega(\ell)$ similar to Eq. (\ref{omegacompleto}) contains products of powers
of $\exp(-
\ell/\xi_l)$ and $\exp(-\ell/\xi_g)$ (see Eq. (\ref{omegas})). 

\subsection{Pure solvent}\label{PS}
We first consider the case of a pure solvent (i.e., $I=0$) near a neutral wall (i.e.,
$\sigma=0$) and at gas-liquid coexistence. For such a system the effective interface
potential in Eq. (\ref{omegacompleto})
reduces to
\be
\beta\omega(\ell)= a_0(T)\exp(- \ell/\xi)+b_0(T)\exp(-2 \ell/\xi)
\ee 
with
\be\label{a}
a_0(T)=(\phi_l-\phi_g)\frac{\chi(T)}{3\xi}\frac{\beta(h_1-g\phi_l)}{\beta g+\frac{\chi(T),
} { 3\xi}}
\ee
and
\be\label{b}
b_0(T)=-(\phi_l-\phi_g)^2\frac{\chi(T)}{12\xi}\frac{\frac{\chi(T)
} { 3\xi}-\beta g}{\beta g+\frac{\chi(T)
} { 3\xi}}.
\ee
For second-order wetting to occur at $T=T_\text{w}$, the coefficient $a_0(T)$ must be negative
for $T<T_\text{w}$, vanish at $T=T_\text{w}$, and be positive for $T>T_\text{w}$. As
$\phi_l>\phi_g$, and because $\phi_l$ can vary only between its value at the triple point
$\phi_l(T_t)$ and
the critical density $\phi_c=\phi_l(T_c)$, $a_0(T)$ fulfills the above mentioned conditions if
\be\label{critical_wetting}
\phi_c<\frac{h_1}{g}<\phi_l(T_t).
\ee
Here and the following we consider $h_1>0$ and $g>0$. The order of the transition is determined by
the higher-order coefficients in the expansion of
$\omega(\ell)$ \cite{Dietrich1988}. If $b_0(T_\text{w})<0$, the transition will be first order while
second-order wetting can occur if $b_0(T_\text{w})>0$. Only in the latter case $a_0(T_\text{w})=0$
determines the wetting transition temperature, so that
\be\label{WT}
\phi_l(T_\text{w})=\frac{h_1}{g}.
\ee
Within the present approach, the wetting transition can be second order if 
\be
\beta g>\frac{\chi(T_\text{w})}{3\xi},
\ee 
and first order if the inequality is reversed. The separatrix between first- and second-order
wetting (i.e., the loci of tricritical wetting \cite{Nakanishi}) is given by
\be\label{separatrix}
\beta g=\frac{\chi(T_\text{w})}{3\xi(T_\text{w})},
\ee
where $\chi(T_\text{w})$ follows from Eqs. (\ref{binodal}) and (\ref{WT}):
\be
\chi(T_\text{w})=\frac{\ln(h_1/g)-\ln(1-h_1/g)}{2h_1/g-1}.
\ee
\subsection{Electrolyte solution}\label{esr}
In the case of an electrolyte solution close to a charged wall the effective interface potential
given by Eq. (\ref{omegacompleto}) has the generic form studied by Aukrust and Hauge \cite{Aukrust}
for a model in
which both the wall-fluid and the fluid-fluid interaction potentials decay exponentially but on
distinct scales.
If we proceed analogously to extract the information about the wetting behavior as before, we
realize that the
electrostatic term $a_I(T)\exp(-2\kappa \ell)$ with
\be\label{aI}
a_I(T)=\frac{4\pi l_B\sigma^2}{\varepsilon_l\kappa}
\ee
has a coefficient which is always positive. (Equation (\ref{omegacompleto}) shows that the 
coefficients $a_0(T)$ (Eq. (\ref{a})) and $b_0(T)$ (Eq. (\ref{b})) do not change upon adding ions.)
Accordingly, the wetting behavior will depend on the
competition between the Debye length $1/\kappa$ and the correlation length $\xi$:
\begin{enumerate}[(i)]
\item \label{1} $1/\kappa<\xi$: In this case the electrostatic term decays faster than the remaining
two
terms in Eq. (\ref{omegacompleto}). Therefore one obtains
the same wetting behavior as for the pure solvent (see Subsec. \ref{PS}).
\item \label{2} $\xi<1/\kappa<2\xi$: In this case the electrostatic term is the dominant subleading
contribution in
the expansion.
Moreover, because $a_I(T)>0$ for all temperatures, the transition can be second order if $a_0(T)$
satisfies the conditions
given by Eq. (\ref{critical_wetting}).
\item \label{3}$1/\kappa>2\xi$: In this case, the electrostatic term is
the leading contribution. As a result, if in the pure solvent the wetting transition is second
order, due to adding ions and due to a nonzero surface charge density at the wall it turns first
order
or the wall becomes wet at all temperatures $T>T_t$.
\end{enumerate}

As discussed in Subsec. \ref{PS}, for the pure solvent it is possible to
determine the separatrix
between
first- and second-order wetting in terms of the surface parameters $h_1$ and $g$ only. Accordingly,
the phase
diagram is of the
type shown in Fig. 2(a) of Ref. \cite{Nakanishi} for $g>0$ and of the type shown there in Fig.
2(b)
for $g=0$. On the other hand, for electrolyte
solutions this separatrix  depends also on the surface charge density, the ionic
strength, and the competition between the Debye and the correlation lengths. As mentioned before our
approach neglects the interaction between ions so that it can be
used only for low ion concentrations, e.g., $I\lesssim10$mM, which corresponds to a
Debye
length $1/\kappa\gtrsim3$nm in water at room temperature. Thus one typically
ends up with case (\ref{3}) ($1/\kappa>2\xi$) except in close proximity to the critical point, where
one can reach case (\ref{2}) ($\xi<1/\kappa<2\xi$) and ultimately case (\ref{1}) ($1/\kappa<\xi$).
Therefore, for $g>0$ the phase diagram for $\sigma\neq0$ is of the type shown in Fig. 2(a)
of Ref. \cite{Nakanishi}, as for the
pure solvent case with $g>0$, but the
separatrix between first- and second-order wetting is shifted closer to the critical point
upon increasing the Debye length, i.e., upon decreasing the ionic strength.

 The wetting behavior will be richer if $\xi_l\neq\xi_g$ (see the discussion below Eq.
(\ref{omegacompleto})). In this case, the possible wetting scenarios will depend on
the competition between the Debye length $1/\kappa$, the correlation length $\xi_g$ of the gas, and
the correlation length $\xi_l$ of the liquid. This creates additional cases compared to the ones
discussed above
(see (\ref{1})-(\ref{3})). Nevertheless, in the present context, far from the critical point
case (\ref{3}) is still the
typical one  with the distinction that here $1/(2\kappa)$ competes
with the maximum of $\xi_l$ and $\xi_g$.

In the limit $\sigma\to0$ one has $a_I(T)\to0$ so that in this case there is no contribution
to the
effective interface potential due to the ions. This is due to the fact that within the present
theory there are no
surfaces fields acting on $\rho_\pm$ if $\sigma=0$. For considering instead
the limit $I\to0$, i.e., $\kappa\to0$, in the expression for $a_I(T)$ one has to use the
saturation value
$|\sigma|=\sigma_{sat}(\kappa\ell)$ given by Eq.
(\ref{limitsigma}), which implies $a_I(T)\sim\kappa^3$. Accordingly, $a_I(T)\to0$ when $\kappa\to0$ 
so that, as expected, in the limit $I\to0$ the pure solvent case is recovered. 

\section{Model with long-ranged interactions}\label{lr}
In this section we consider systems in which the solvent exhibits attractive long-ranged interaction
potentials among the solvent particles as well as between the
wall and the solvent particles. As in the previous section, we are interested in an analytic
expression
for the effective interface potential $\omega(\ell)$. Following Ref.~\cite{Getta1998} we model the
attractive part of the pair potential between the solvent particles, as it enters the density
functional, as
\be\label{wr}
\bar w(r)=\frac{A_f}{(1+r^2)^3}
\ee
with $A_f<0$ and the substrate potential as
\be\label{Vz}
V(z>0)=-\sum_{i\ge3}\frac{u_i}{z^i}
\ee
with $u_3>0$ corresponding to an asymptotically attractive interaction. The contribution
$\sim u_4$ is generated, inter alia, by the discrete lattice structure of the substrate or by
a thin overlayer
\cite{Dietrich1988} and thus it can be tuned. The substrate potential $V(z)$ diverges
for $z\to0$. Therefore the solvent density $\phi(z)$ must vanish for $z\to0$. In the following
this effect is taken into account approximately by replacing the short-ranged part of $V(z)$
in Eq. (\ref{Vz}) by a
hard-wall potential positioned at $z=d_w$; the distances $z$ are still measured from $z=0$.
(Beyond this sharp-kink approximation for the wall-liquid interfacial profile, $d_w$ is replaced by
the moment $d_{wl}^{(1)}$ (Eq. \ref{dwl1}) of the profile $\phi_{wl}(z)$.) This implies that in
the
present section the short-ranged
description of the surface-fluid interaction given in terms of the surface parameters $h_1$ and $g$
in the previous section has to be shifted from $z=0$ to $z=d_w$ (see Eq. (\ref{functional})). On the
other hand, in order to account for the long-ranged attractive part of $V(z)$ (i.e., for $z\gg
d_w$), here only the first two terms of the sum in Eq. (\ref{Vz}) are considered. The
functional
form in Eq. (\ref{wr}) facilitates to carry out subsequent integrals analytically. 
These long-ranged interactions are treated as a perturbation of the grand canonical functional
in Eq.
(\ref{functional}):

\be\label{functional2}
\Omega[\phi(\mathbf r),\rho_\pm(\mathbf r)]=\Omega_0[\phi(\mathbf r),\rho_\pm(\mathbf
r)]+\Delta\Omega[\phi(\mathbf r)]
\ee
where $\Omega_0[\phi(\mathbf r),\rho_\pm(\mathbf r)]$ is given by Eq. (\ref{functional}) and
\be\label{DeltaOmega}
\begin{aligned}
\Delta\Omega[\phi(\mathbf r)]&=\frac{1}{2}\int d^3r\int
d^3r'\bar w(|\mathbf{r}-\mathbf{r'}|)\phi(\mathbf{r})\phi(\mathbf{r})\\
&+\int d^3r\rho_wV(\mathbf{r})\phi(\mathbf{r}).
\end{aligned}
\ee
The integrations run over the half space $\{\mathbf r=(x,y,z\ge d_w)\}$, $\bar w(r)$ is given by Eq.
(\ref{wr}), and $V(\mathbf r)$ is given by Eq. (\ref{Vz}); $\rho_w$ is the particle number
density of
the substrate.
Concerning the interaction between the
solvent
particles, it turns out that it is most suitable captured by the quantity \cite{Dietrich1988},
\be
t(z):=\int_z^\infty dz'\int d^2 r'_{||}\bar w\left(\left({r'}_{||}^2+{z'}^{2}\right)^{1/2}\right).
\ee
For large distances and non-retarded van der Waals forces one has 
\be
t(z\to \infty)=-\left(\frac{t_3}{z^3}+\frac{t_4}{z^4}+\cdots\right),
\ee
which defines the coefficients $t_3>0$ and $t_4$. For the present model this implies
\be\label{t3}
t_3=-\frac{\pi A_f}{6},
\ee
\be\label{t4}
t_4=0.
\ee

The addition of the long-ranged pair potential between solvent particles (Eq. (\ref{wr})) modifies
the bulk grand canonical potential per $k_BT$ of the pure solvent (i.e., $\rho_\pm=0$) (see Eq.
(\ref{omegabulk})). Accordingly, in this system the bulk densities $\phi_l$ and $\phi_g$
minimize the modified
bulk grand canonical potential density given by
 
\be\label{omegabulk_lr}
\begin{aligned}
\beta\Omega_{b,lr}(\phi,\rho=0)&=\phi(\ln(\phi)-\beta\tilde\mu_{\phi}
)+(1-\phi)\ln(1-\phi)\\
&+\tilde\chi(T)\phi(1-\phi)
\end{aligned}
\ee
with the shifted solvent chemical potential $\tilde\mu_\phi=\mu_\phi-\pi^2A_f/8$ and the modified
Flory-Huggins parameter $\tilde{\chi}(T)=\chi_S+\chi_H/T-\pi^2\beta A_f/8$, i.e.,
the RPA-like perturbation $\Delta\Omega[\phi(\mathbf{r})]$ in Eq.~(\ref{DeltaOmega}) changes
only 
the enthalpic part of the Flory-Huggins parameter.
The binodal $T_{bi,lr}(\phi)$ is again of the form given in Eq.~(\ref{binodal}) but with
$\chi(T)$ replaced
by $\tilde\chi(T)$.
Hence the critical point is located at $(\phi_c=1/2$, $\tilde\chi(T_c)=2)$, i.e.,
\be\label{Tclr}
T_c=\frac{\chi_H-\pi^2A_f/(8k_B)}{2-\chi_S}.
\ee
The bulk correlation length is now given by (see Appendix \ref{axi})
\be
\label{chilr}
\frac{1}{\xi^2}=
\frac{\dps\frac{3}{\tilde\chi(T)}\Big(\frac{1}{\phi}+\frac{1}{1-\phi}-2\tilde\chi(T)\Big)}
     {\dps 1 + \frac{\pi^2\beta A_f}{8\tilde\chi(T)}}.
\ee

In a first-order perturbative theory approach (see Appendix \ref{1PT})
the influence of
$\Delta\Omega[\phi(\mathbf r)]$ on the wetting behavior of the electrolyte solution can be
determined by inserting into Eq. (\ref{functional2}) the solutions  $\phi^{(0)}(\mathbf
r)$
and $\rho^{(0)}_\pm(\mathbf r)$ as obtained from $\Omega_0[\phi(\mathbf r),\rho_\pm(\mathbf
r)]$ (see Sec. \ref{sr}). The superscript $(0)$ denotes these solutions as the ones obtained from
the
unperturbed functional $\Omega_0$. 

Following the same procedure as described in Sec. \ref{sr}, i.e., expanding the local part of
the grand canonical functional in Eq. (\ref{functional2}) around the sharp-kink density profiles in
Eqs. (\ref{sharpkinkphi}) and (\ref{sharpkinkI}), for $\ell\to\infty$ we obtain the
following
form for the
effective interface potential (see Appendix  \ref{integrals}):
\be\label{omega2}
\begin{aligned}
\beta\omega(\ell\to \infty)&\simeq\frac{a_1(T)}{\ell^2}+\frac{b_1(T)}{\ell^3}+\cdots \\
&+a_0(T)\exp(-\ell/\xi)+b_0(T)\exp(-2\ell/\xi)\\
&+a_I(T)\exp(-2\kappa \ell),
\end{aligned}
\ee
where ellipses stand for further subdominant terms as powers of $1/\ell$. As in the absence
of long-ranged interactions the ions enter into $\omega(\ell)$ only via the last term. The analytic
expressions for the
coefficients $a_1(T)$ and $b_1(T)$ are given in Appendix \ref{clr}, $a_0(T)$ and $b_0(T)$ are given
by
Eqs. (\ref{a}) and (\ref{b}), respectively, and $a_I(T)$ is
given by Eq. (\ref{aI}). Corrections to the coefficients $a_0(T)$ and $b_0(T)$ due to the
long-ranged interactions (Eqs. (\ref{wr}) and (\ref{Vz})) are
neglected because these long-ranged interactions are treated as a small perturbation to the model
with short-ranged interactions only. 
The sign of the coefficients $a_1(T)$, $b_1(T)$, $a_0(T)$, and $b_0(T)$ can change with $T$ while
$a_I(T)$ is always positive (see Appendix \ref{clr}).
As discussed for short-ranged interactions in the previous Sec.~\ref{sr}, the order of the
wetting transition can be inferred from the analysis of these coefficients. 
They depend on seven parameters: $\chi_S$, $\chi_H$, $A_f$, $u_3\rho_w$, $u_4\rho_w$,
$h_1$, and $g$.
The value of $\chi_S$ is typically much smaller than unity \cite{Rubinstein} so that we set 
$\chi_S=0$ in the following.
Moreover, in the discussion below, the amplitude $A_f$ is
chosen to be in the range $(0.04-0.15)\times10^{-19}$~J, which corresponds to typical
strengths of the
van der Waals interaction in condensed phases (see Ref.~\cite{Israelachvili}) and $\chi_H$
is determined via of
Eq.~(\ref{Tclr}) using the critical temperature $T_c=647$~K of water.
Finally, $u_3\rho_w$ is fixed by specifying the temperature $T^{(a_1)}$ at which $a_1(T^{(a_1)})=0$
given by the implicit relation (see Eq. \ref{D1}) $\phi_l(T^{(a_1)})=u_3\rho_w/t_3=-6u_3\rho_w/(\pi
A_f)$
($u_3>0$, $A_f<0$); in the
case of a critical wetting transition this temperature coincides with the wetting
transition temperature $T_w=T^{(a_1)}$.
With these choices the only remaining free parameters in the following analysis are
 $u_4\rho_w$, $h_1$, and
$g$. However,
their values are constrained by the condition $b_1(T_w)>0$ for critical wetting (see
Eq.~(\ref{D2})).
Due to the additional presence of the  parameters $A_f$,  $u_3\rho_w$, and $u_4\rho_w$  of the
long-ranged
interactions, in that case the corresponding discussion is
slightly more difficult than the one for  short-ranged interactions only as studied in the
previous section. 

We start this discussion by analyzing the pure solvent case, i.e., $a_I=0$. In this case, the
necessary
conditions for the occurrence of critical wetting are (Eq. (\ref{omega2}) and Ref.
\cite{Dietrich1988})
\be
 a_1(T_{\text w})=0,\ \ a_1(T<T_{\text w})<0,\ \text{and}\ b_1(T_w)>0
\ee 
i.e., $T^{(a_1)}=T_w$, and, as before, one obtains conditions for the parameters of the pair
potentials  (see Eqs. (\ref{D1}) and (\ref{D2})): 
\be\label{a1c}
\begin{aligned}
\phi_c/\rho_w&<u_3/t_3<\phi_l(T_t)/\rho_w
\end{aligned}
\ee
and
\be\label{b1c}
\rho_wu_4-3t_3\phi_l(T_w)\left[d_w+d_{wl}^{(1)}\right]>0,
\ee
with $d_{wl}^{(1)}$ given by Eq. (\ref{dwl1}) and $\phi_l(T_w)=u_3\rho_w/t_3$.
 
Although necessary, these conditions are not sufficient for critical wetting to occur. Large
negative values of the coefficient $a_0(T)$ of the exponentially decaying contribution
can still lead to a first-order
wetting transition  even if $b_1(T^{(a_1)})>0$.  Within the present model one has $a_0(T)>0$
for
$h_1/g>\phi_l(T)$ (see Eq. (\ref{a})). If $b_1(T^{(a_1)})<0$ the
wetting transition is always first order. However, in the case of
a first-order wetting transition all details of
$\omega(l)$, and not only its leading contributions, matter for a reliable
description of the character
of the transition and for determining the corresponding wetting transition
temperature. Hence, an
asymptotic expansion of $\omega(\ell)$ as in Eq. (\ref{omega2}) is not conclusive in the case of
first-order wetting.

For wetting of a wall by a one-component fluid with short- and long-ranged interactions and
based on a Cahn type theory, in Refs. \cite{Indekeu1999, Indekeu2000} a wetting scenario has been
predicted which involves a succession of two interfacial phase transitions upon increasing
$T$. The first of these two transitions is a discontinuous jump between two
finite values $\ell_1$ and $\ell_2>\ell_1$ of the film
thickness $\ell$ at two-phase coexistence and is referred to as a ``thin-thick transition''. The
second one is the standard second-order wetting transition at $T=T_w$. (In Refs.
\cite{Indekeu1999, Indekeu2000} the possibility of a thin-thick transition preceding a first-order
wetting transition has not been discussed). This wetting scenario can
be explained  in terms of the competition between the
short- and long-ranged interactions. Such a thin-thick transition precedes the critical
wetting transition only if the short-ranged interactions would give rise to a first-order wetting
transition in the case that the long-ranged interactions were negligible. Because the present
theory involves both short- and long-ranged interactions, the occurrence of this wetting
scenario can be
checked for the pure solvent case. In this case, the separatrix between first- and second-order
wetting
is given by Eq. (\ref{separatrix})  for the model with short-ranged interactions only (e.g.,
for $g=1$ the transition will be first order in the pure
solvent case without long-ranged interaction if $h_1>0.49$). By choosing a proper set of parameters
(see the discussion above) we have been able to observe the occurrence of this two-stage
transition for the pure
solvent within our model
for $\pi^2A_f\lesssim0.55\times10^{-19}$~J, $\phi_l(T_w)=u_3\rho_w/t_3=0.7$, $u_4=2.3\times t_3$,
$g=1$, and $h_1=0.76$, such that the condition for second-order wetting given by Eq.
(\ref{b1c}) is satisfied.

This thin-thick transition has also been
observed for wetting of a wall by a one-component fluid in models with short-ranged
interactions only \cite{Piasecki,Indekeu,Langie} and with long-ranged interactions only
\cite{Dietrich1985}. Furthermore it has been observed experimentally for wetting of hexane on
water
\cite{Shahidzadeh}. In Ref.
\cite{Piasecki} this thin-thick transition has been
observed for a
generalization of the Sullivan model \cite{Sullivan}, in which in addition to the exponentially
decaying wall-fluid potential a square-well attraction has been
included. A thin-thick transition was also analyzed in Ref. \cite{Indekeu} for a
Landau theory of wetting which includes an extra surface term $h_3\left(\phi(0)\right)^3$ linked to
the substrate
potential (see Ref. \cite{Nakanishi} and Eq. (\ref{functional})). In Ref. \cite{Langie} it has been
shown that the behavior of the model in Ref. \cite{Piasecki} can be mapped onto that used in
Ref. \cite{Indekeu}. With that it turns out that the thin-thick transition
predicted in
Refs. \cite{Piasecki} and \cite{Indekeu} involves short-ranged forces only and is due to
the competition between two opposing (effective) surface fields at the same surface, one
favoring wetting and the other favoring drying. Such a competition between surface fields is
not
considered here. Therefore within our model a thin-thick transition does not occur
in the
pure solvent case with short-ranged interactions only (see Sec. \ref{sr}). 

The influence of ions and of surface charges on the wetting
behavior of electrolytes with solvents governed by short- and long-ranged forces
differs qualitatively from the one
discussed in Subsec. \ref{esr}, because in this case the leading contributions to
$\omega(\ell\to
\infty)$ decay algebraically as function of the film thickness $\ell$. Accordingly, the
contribution
due to the ions and the charged wall can enter at most as the leading non-algebraic term in the
expansion for $\ell\to\infty$; this is the case if the Debye length $1/\kappa$ is larger than
(twice) the 
bulk correlation length $\xi$ (see Subsec \ref{esr}). 

\begin{figure*}[!htb]
\includegraphics[width=12cm]{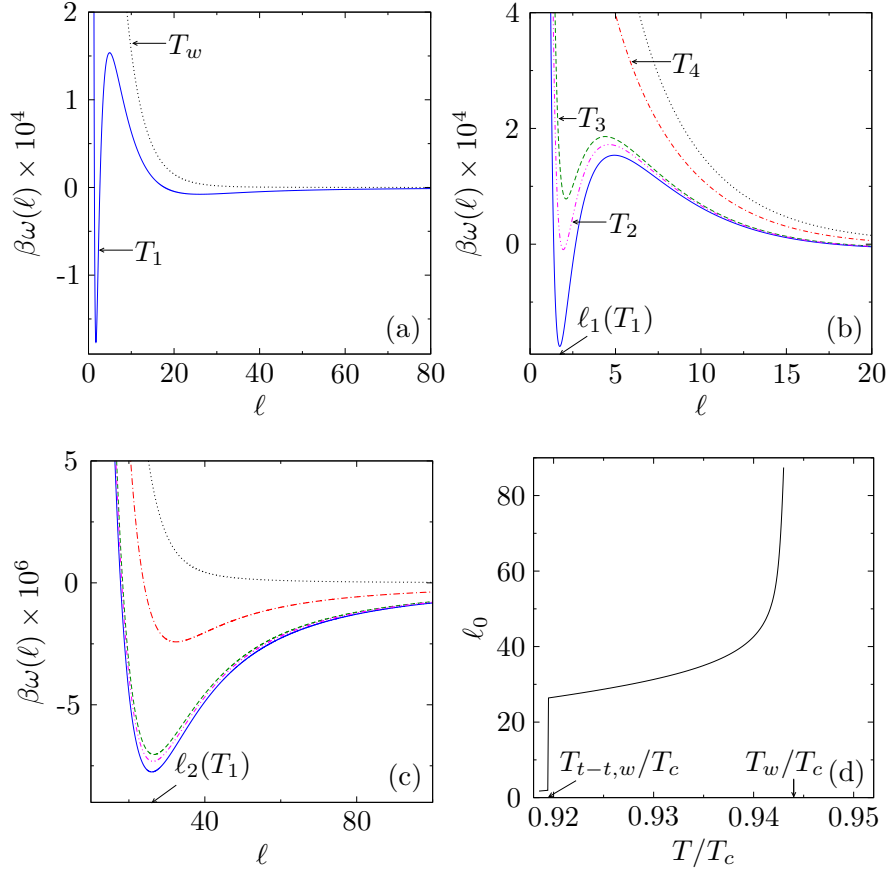}
   \caption{Effective interface potential $\omega(\ell)$ for systems governed by short- and
long-ranged interactions as function of the thickness $\ell$ of
the liquid film at gas-liquid coexistence in the presence of ions for the case  that the pure,
i.e., salt-free,
solvent exhibits a critical wetting transition (without being preceded by a thin-thick
transition).
The parameters used are $A_f/k_B=-1013$K, $u_3=0.7\times t_3$ (i.e.,
$\phi_l(T^{(a_1)}=T_w)=0.7$), $u_4=2.28\times t_3$,
$g=1$, $h_1=0.76\times g$, $I=1mM$, and 
$\sigma=0.1\mu$C/cm$^2$ (see main text).
The effective interface potential $\omega(\ell)$ has two local minima  (at $\ell_1(T)$ (see
 (a) and (b))
and
$\ell_2(T)$ (see (c))
with $\ell_1<\ell_2<\infty$), one of the two being the global one at a given temperature
(see (a)).
They have the
same
depth
at $T=T_{t-t,w}\approx0.919\times T_c$ (not apparently visible). For $T>T_{t-t,w}$ the film
thickness $\ell_2(T)$ is the
global minimum and diverges continuously $~1/(T_w-T)$ as $T\to T_w\approx0.944\times T_c$ (see
(c)). The global
minimum
$l_0(T)$ as a function of temperature is plotted in (d). At $T_{t-t,w}$ the film thickness exhibits
a finite jump and subsequently diverges smoothly for $T\nearrow T_w$.
Accordingly, the system undergoes a thin-thick wetting transition at $T_{t-t,w}$, followed by a continuous one at $T_w$. 
Five different temperatures,  $T_1\approx0.918\times T_c$, $T_2\approx0.919\times T_c$,
$T_3\approx0.92\times T_c$,
$T_4=0.932\times T_c$
and $T_w$ are displayed in (a), (b), and (c) (using a common color code) with $T_1<T_2\lesssim
T_{t-t,w}<T_3<T_4<T_w$. (Note the different scales of the axes.)
The film thickness $\ell$ is measured in units of $a$ such that $a^3$ is the volume of a solvent
particle. Densities are measured in units of $a^3$.}
\label{thin-thick}
\end{figure*}

We have considered various parameter sets  $(h_1,g,u_4,T^{(a_1)})$  chosen such
that the pure solvent with short- and long-ranged interactions near a charge neutral wall
(i.e., for $a_I(T)=0$) exhibits a second-order wetting transition  at $T_w(I=0,\sigma=0)$
\emph{without} being preceded by a thin-thick transition (i.e., different from the above
scenario).
 For fixed ionic strength $I\neq0$ and upon increasing the surface charge density
$\sigma$, due to $a_I(T)\sim\sigma^2/\sqrt{I}$ (Eq. \ref{aI}) $\omega(\ell)$ rises at finite
film thickness $\ell$ to the effect that the
wetting transition temperature $T_w(I,\sigma)$ decreases for increasing surface charge density
$\sigma$ \cite{Ibagon}. Moreover, for fixed surface charge density $\sigma$ the
wetting transition temperature $T_w(I,\sigma)$ decreases upon decreasing 
the ionic strength $I$ (i.e., increasing the amplitude $\sigma^2/\sqrt{I}$ and the Debye
length $1/\kappa\sim1/\sqrt{I}$) \cite{Ibagon}. In addition, the positive and monotonically
decreasing (as a function of increasing
$\ell$) contribution
$a_I(T)\exp(-2\kappa \ell)$ to $\omega(\ell)$
does lead to a thin-thick transition preceding the critical wetting transition which is absent
without ions.
Figure~\ref{thin-thick}  shows the curves for $\omega(\ell)$ corresponding to the temperatures
$T_1=0.918\times T_c$, $T_2=0.919\times T_c$, $T_3=0.92\times T_c$, $T_4=0.932\times T_c$, and
$T_w=0.944\times T_c$ with
$T_1<T_2\lesssim T_{t-t,w}<T_3<T_4<T_w$, i.e., the \textit{t}hin-\textit{t}hick transition occurs in
between the
temperatures $T_2$
and $T_3$, whereas
the critical wetting transition takes place at the wetting temperature $T_w$.

However,  in the case that the pure solvent exhibits a second-order wetting transition, which
is preceded by a
thin-thick wetting 
transition, the effect of the term due to the ions and to the surface charge density
($a_I(T)\neq0$), in the case
$1/\kappa>2\xi$, is to decrease the thin-thick wetting transition temperature $T_{t-t,w}$ and to
increase the value of the jump in film thickness.

The case of $a_I(T)\neq0$ for a system in which a pure solvent with short- and
long-ranged interactions near a charge neutral
wall exhibits a first-order wetting transition is not
discussed here, because  within the present approach only the leading
contributions of the effective interface potential for $\ell\to\infty$ are analytically
accessible
(see Eq. (\ref{omega2})) and
reliable knowledge of the behavior of $\omega(\ell)$ for small $\ell$, which is
particularly important for
first-order wetting transitions, is lacking. Therefore, in order to
be able to analyze the effect of the ions and of the surface charge density on 
solvents which without ions exhibit first-order wetting transitions, more details of the
effective interface potential are
needed.

The thin-thick wetting transition at two-phase
coexistence, which precedes a standard second-order wetting scenario, has been discussed in the
context of wetting
in electrolytes in Ref.
\cite{Denesyuk2004} for a model of an ionic solution close to a charged wall in which the
solvent-solvent and solvent-wall interactions are short-ranged only and the contribution of the
ions to
the effective interface potential is calculated by solving the full Poisson-Boltzmann equation
instead of the linearized one as in the present study (see Sec. \ref{sr}). The thin-thick
transition
in Ref. \cite{Denesyuk2004} occurs in a restricted region of the parameter space, provided
that the transition in the pure solvent is first order and that $1/\kappa<2\xi$, i.e, for
large ionic strength.

In contrast, within the present
approach the \textit{combined} presence of short- \textit{and} long-ranged interactions is
taken into account. As discussed above for the case of a
pure
solvent with short- \textit{and} long-ranged interactions, a thin-thick transition
will
precede a long-ranged critical
wetting transition only if the short-ranged interactions alone would give rise to a
first-order wetting
transition in the case that the long-ranged interactions were negligible \cite{Indekeu1999,
Indekeu2000}. This is precisely the case we encounter in the present context for the electrolyte
solution when solvent-solvent and solvent-wall long-ranged interactions are taken into account: In
the absence of these long-ranged interactions the transition is first-order if
$1/\kappa>2\xi$ (see Subsec. \ref{esr}), such that  $\ell$ jumps from $\ell_1$ to
$\ell_2=\infty$
(see Fig.
\ref{thin-thick}). Once the long-ranged interactions are taken into
account they block the jump of $\ell$ to $\ell_2=\infty$ and limit this jump to one with a
finite value $\ell_2<\infty$.  Once $\ell$ has reached the value $\ell_2$ a further increase in
temperature leads to the unfolding of the standard wetting scenario under the aegis of long-ranged
interactions at $T_w>T_{t-t,w}$. Therefore, the thin-thick
wetting transition is the remnant of the first-order wetting transition that would occur in
the
electrolyte solution  if the long-ranged solvent-solvent and solvent-wall interactions  were
negligible (see Subsec. \ref{esr}).

We briefly give the main points of the literature and of our results.
\begin{enumerate}[(I)]
 \item If in the pure solvent short-ranged interactions favor first-order wetting but
additional
long-ranged interactions produce second-order wetting, one finds a thin-thick transition followed
by the continuous wetting transition \cite{Indekeu1999, Indekeu2000}. We confirmed this behavior
for our model, which allows for the thin-thick transition only via the competition between
short- and long-ranged interactions.
\item A thin-thick transition can be observed in the pure solvent even if there are only
short-ranged \cite{Piasecki, Indekeu, Langie} or only long-ranged \cite{Dietrich1985}
interactions.
\item We have considered a solvent with short- \textit{and} long-ranged interactions which exhibits
a
second-order wetting transition without being preceded by a thin-thick transition. Adding ions
renders such a short-ranged contribution to the effective interface potential that the resulting
effective short-ranged interactions favor first-order wetting. This leads to a thin-thick
transition
preceding the continuous long-range type wetting transition. This mechanism is analogous to the one
in (I).
\item If the solvent with short- and long-ranged interactions undergoes a continuous wetting
transition, which \textit{is} preceded by a thin-thick transition, adding ions decreases the
transition temperature of the latter and increases the jump in film thickness.
\item If the pure solvent is governed by short-ranged interactions only and exhibits a
first-order
wetting transition, adding ions can lead to a continuous wetting transition preceded by a thin-thick
transition, provided that $1/\kappa<2\xi$. 
\item If the solvent is governed by short-ranged interactions only, adding ions renders a
first-order wetting transition for $1/\kappa>2\xi$. Adding further long-ranged interactions, which
favor continuous wetting, renders a second-order wetting transition of the long-range type,
preceded by a thin-thick transition.
\end{enumerate}


\section{Conclusions and Summary}\label{S}

We have implemented an improvement over the approximation of step-like varying density profiles in
order to derive analytic
expressions for the effective interface potential $\omega(\ell)$ of electrolyte solutions near
charged walls. This approach consists of performing a Taylor expansion up to second order of the
local part of the grand canonical density functional around piecewise constant density profiles of
the solvent and of the ions. The
resulting mean-field
expressions for the effective interface potential allow one to predict general trends for the
wetting behavior of electrolyte 
solutions in terms of the relevant system parameters such as the ionic strength and the surface
charge 
density.

The present analysis, which is valid in the case of low ion density $I$, shows that in the case of
short-ranged
solvent-solvent and solvent-wall interactions
wetting transitions in the presence of 
electrostatic interactions are typically first order. This result can be
explained in terms of the competition between the two characteristic length scales in the system,
i.e., the bulk correlation length $\xi$ in the wetting liquid phase and the Debye length $1/\kappa$.
If $1/\kappa>2\xi$, which is typically the case for dilute electrolyte solutions away from
(bulk)
critical points, a wetting transition at two-phase coexistence will be always first order
irrespective of its order in the pure, i.e., salt-free, solvent.
First-order wetting transitions in electrolyte solutions with solvent
interactions being short-ranged only have been observed in
previous
studies, too \cite{Denesyuk2003, Denesyuk2004, Oleksy2009, Oleksy2010, Ibagon}. It is the merit
of
the present analysis of the effective interface potential to provide a transparent rationale
for the pre-eminence of first-order wetting in electrolyte solutions in terms of competing length
scales.
Moreover, if in those systems in addition long-ranged
solvent-solvent and solvent-wall interactions,  which favor a critical wetting transition,
are present,
our analysis reveals the possibility of a wetting scenario which actually corresponds to
a sequence of two wetting transitions:
first an electrostatically induced (i.e., $1/\kappa>2\xi$) discontinuous jump between two
finite wetting film
thicknesses which upon raising the
temperature is
followed by a
continuous divergence of the wetting film thickness $\ell$ (see Fig. \ref{thin-thick}).

Within the present approach, in the case of short-ranged interactions the analytic expressions for
the coefficients of the  exponential terms in the effective interface
potential $\omega(\ell)$ (Eq. (\ref{omegacompleto})) are simple and the necessary
conditions for first- and second-order wetting can be translated explicitly into conditions for
the parameters of the interaction potentials. However, in the case of additional long-ranged
interactions extra parameters make such a kind of translation more difficult. Nevertheless,
by choosing a set of parameters for the
interaction potentials based on actual values for the Hamaker constant and the critical
temperature and by fulfilling corresponding necessary conditions for the occurrence of
second-order
wetting in the pure solvent (formulated in terms of the coefficients $a_1(T)$ and
$b_1(T)$ (Eqs. (\ref{omega2}),
(\ref{D1}), and
(\ref{D2}))), we have been able to analyze
the effect of ions and of the surface charge density of the confining wall on the wetting
behavior. We have found that if the pure solvent exhibits a second-order wetting transition
governed asymptotically by long-ranged interactions, adding ions typically introduces a
thin-thick
transition which precedes the  ultimate continuous wetting transition (see Fig.
\ref{thin-thick}).
 We have
been able to put the occurrence of such a thin-thick wetting transition at gas-liquid coexistence
into the context of the literature, which discusses such a transition as the result of the
interplay between short- and long-ranged interactions. Here the corresponding short-ranged effective
interactions relevant for that are provided by the ions.

\appendix
\section{Double parabola approximation for the pure solvent}\label{DPS}
The double parabola approximation (DPA) has been widely used in the context of wetting
phenomena in order to obtain analytically tractable density functionals \cite{Meister1983,
Lipowsky1984, 
Holyst1987, Gelfand1987, Jin1993, Iwamatsu1993, Parry2006, Wojtowicz2013}.
Within this approximations, the grand canonical functional for the pure solvent,
i.e., $\rho_\pm=0$ and $\mathbf
D(\mathbf r,[\rho_\pm])=0$ (see Eq.~(\ref{functional})) is given by
\begin{equation}\label{DP}
\begin{aligned}
\frac{\beta\Omega_{DPA}[\phi(z)]}{A}&=\int
dz\biggl\{F_{DPA}(\phi(z))+\frac{\chi(T)}{6}
\left(\frac{d\phi(z)}{dz}\right)^2\\
&-\phi(z)\beta\mu_\phi\biggr\}\\
&-\beta
h_1\phi(z=0)+\beta\frac{g}{2}\left(\phi(z=0)\right)^2
\end{aligned}
 \end{equation}
with
\be\label{FDP}
F_{DPA}(\phi)=C(T)
\begin{cases}
(\phi\!-\!\phi_l(T))^2, & \phi\!>\!\frac12(\phi_l(T)\!+\!\phi_g(T))\\
(\phi\!-\!\phi_g(T))^2, & \phi\!<\!\frac12(\phi_l(T)\!+\!\phi_g(T)),
\end{cases}
\ee
where $\phi_l(T)$ and $\phi_g(T)$ are, respectively, the (temperature dependent) liquid and gas
bulk densities at coexistence, and $C(T)$ is fixed later in order to render the bulk
correlation length. Upon construction, the DPA requires an underlying theory which provides
expressions for the bulk densities
and the curvature of the local free energy density at coexistence \cite{Meister1983, Lipowsky1984, Holyst1987,
Gelfand1987, Jin1993, Iwamatsu1993, Parry2006, Wojtowicz2013}.
For brevity we shall refrain from indicating the temperature dependence  in our notation.
Within this approach, for a given profile $\phi(z)$ the assigned film thickness $\ell_{DPA}$ is
defined as
\be
\phi(z=\ell_{DPA})=\frac12(\phi_l+\phi_g).
\ee
Minimization of the functional in
Eq. (\ref{DP}) leads to 
\begin{widetext}
\be\label{EqDP1}
\phi''_{DPA}(z)=
\begin{cases}
\frac{6C}{\chi}(\phi_{DPA}(z)-\phi_l)-\beta\mu_\phi, & \phi_{DPA}(z)>\frac12(\phi_l+\phi_g)\\
\frac{6C}{\chi}(\phi_{DPA}(z)-\phi_g)-\beta\mu_\phi, & \phi_{DPA}(z)<\frac12(\phi_l+\phi_g)
\end{cases}
\ee
\end{widetext}
with the boundary conditions 
\be\label{BCDP}
\begin{aligned}
\frac{\chi}{3}\phi'_{DPA}(0)&=\beta (-h_1+g\phi_{DPA}(0)),\\
\phi_{DPA}(\ell_{DPA})&=\frac12(\phi_l+\phi_g),\\
\phi_{DPA}(\infty)&=\phi_g.
\end{aligned}
\ee
At two-phase coexistence $\mu_\phi=0$ (as for the pure solvent case, i.e., $\rho_\pm=0$, in the
present model given by Eq.~(\ref{omegabulk})) and Eq. (\ref{EqDP1}) reduces to
\be\label{EqDP}
\phi''_{DPA}(z)=
\begin{cases}
\frac{6C}{\chi}(\phi_{DPA}(z)\!-\!\phi_l), & \phi_{DPA}(z)\!>\!\frac12(\phi_l\!+\!\phi_g)\\
\frac{6C}{\chi}(\phi_{DPA}(z)\!-\!\phi_g), & \phi_{DPA}(z)\!<\!\frac12(\phi_l\!+\!\phi_g).
\end{cases}
\ee
Comparison with Eq.~(\ref{eqphi}) leads to $C=\chi/(6\xi^2)$.
Equation (\ref{EqDP}) together with the boundary conditions in Eq. (\ref{BCDP}) yields 
\be\label{SDP}
\phi_{DPA}(z)\!=\!
\begin{cases}
C_1\exp\left(\!\frac{z}{\xi}\!\right)\!+C_2\exp\left(\!-\frac{z}{\xi}\right)\!+\phi_l, &\!\!\!\!
0\!\leq\!z\!\leq\!\ell_\text{DPA}\\
C_3\exp\left(\!-\frac{z}{\xi}\right)+\phi_g, &\!\!\!\! z\!\geq\! \ell_\text{DPA}
\end{cases}
\ee
where
\be
\begin{aligned}
C_1&=\frac{\frac{\phi_g\!-\!\phi_l}{2}\left(\beta
g\!+\!\frac{\chi}{3\xi}\right)\!-\!\beta(h_1\!-\!g\phi_l)\exp(-
\ell_\text{DPA}/\xi)}{\left(\beta
g\!+\!\frac{\chi}{3\xi}\right)\exp(\ell_\text{DPA}/\xi)\!+\!\left(\frac{\chi}{3\xi}\!-\!\beta
g\right)\exp(- \ell_\text{DPA}/\xi)},\\
C_2&=\frac{\frac{\phi_g\!-\!\phi_l}{2}\left(\frac{\chi}{3\xi}\!-\!\beta
g\right)\!+\!\beta(h_1\!-\!g\phi_l)\exp(\ell_\text{DPA}/\xi)}{\left(\beta
g\!+\!\frac{\chi}{3\xi}\right)\exp(\ell_\text{DPA}/\xi)\!+\!\left(\frac{\chi}{3\xi}\!-\!\beta
g\right)\exp(-
\ell_\text{DPA}/\xi)},\\
C_3&=\frac{\phi_l-\phi_g}{2}\exp\left(\ell_\text{DPA}/\xi \right).
\end{aligned}
\label{DPcoeff}
\ee
The comparison between Eqs. (\ref{solutionphi}) and (\ref{SDP}) shows that the
coefficients $\hat A_l$, $\hat B_l$, and $\hat B_g$ there play the same role as the coefficients
$C_1$, $C_2$, and $C_3$, respectively, here. At coexistence and for
$\ell=\ell_\text{DPA}$ in Eq.~(\ref{consts_co}) one obtains
\be
\begin{aligned}
   C_1 &= A_{l}(1 + \mathcal{O}(\exp(-\ell_\text{DPA}/\xi))), \\
   C_2 &= B_{l}(1 + \mathcal{O}(\exp(-\ell_\text{DPA}/\xi))), \\
   C_3 &= B_{g}(1 + \mathcal{O}(\exp(-\ell_\text{DPA}/\xi))), 
\end{aligned}
\label{relativedifference}
\ee
i.e., the relative difference between the coefficients of the profiles in Eq.~(\ref{solutionphi})
and in
Eq.~(\ref{SDP}) is exponentially small for film thicknesses
$\ell=\ell_\text{DPA}\gg\xi$ large compared to the
bulk correlation length.

\section{Bulk correlation length of the pure solvent}\label{axi}
In the case of a bulk \textit{p}ure solvent ($I=0$), the density functional
$\Omega_0$ given by Eq. (\ref{functional})
reduces to

\be\label{solvent}
\begin{aligned}
\beta\Omega_0^{(p)}[\phi(\mathbf{r})]&=\int\! d^3r\biggl\{\phi(\mathbf
r)(\ln(\phi(\mathbf r))-\beta\mu_{\phi}
)\\
&+(1-\phi(\mathbf r))\ln(1-\phi (\mathbf r))\\
&+\chi(T)\phi(\mathbf r)(1-\phi(\mathbf
r))+\frac{\chi(T)}{6}\left(\nabla\phi(\mathbf r)\right)^2\biggr\}.
\end{aligned}
\ee
If we consider a spatially uniform equilibrium state $\phi$, the corresponding two-point correlation
function $G(r)=\phi^2h(r)=\phi^2(g(r)-1)$ is obtained from $G(r)= G(\mathbf{r},\mathbf{0})$, with
the inverse $G^{-1}({\mathbf r}, {\mathbf r'})=\frac{\delta^2\Omega}{\delta \phi(\mathbf r)\delta
\phi(\mathbf r')}$, where $\int d^3r''G(\mathbf r,\mathbf r'')G^{-1}(\mathbf r'',\mathbf
r')=\delta(\mathbf
r-\mathbf r')$ \cite{Evans}. From Eq. (\ref{solvent}) one obtains
\begin{multline}\label{B2}
G^{-1}(\mathbf r, \mathbf r')=\delta(\mathbf r- \mathbf
r')\left(\frac{1}{\phi}+\frac{1}{1-\phi}-2\chi(T)\right)\\-\frac{\chi(T)}{3}\nabla^2\delta(\mathbf
r-
\mathbf r').
\end{multline}
The corresponding Fourier transform, $\hat G^{-1}(k)=\int
d^3r\,G^{-1}(\mathbf{r},\mathbf{r}' )e^{-i\mathbf k\cdot(\mathbf r - \mathbf r')} $
with dimensionless $\mathbf k$, is given by
\be
\hat{G}^{-1}(k)=\frac{\chi(T)}{3}\left(\frac{1}{\xi^2}+k^2\right),
\ee
where Eq. (\ref{xi}) has been used. $\hat G(k)$, which is proportional
to the
static structure factor, can be written in the form
\be
\hat
G(k)=\frac{\frac{3\xi^2}{\chi(T)}}{1+(k\xi)^2}=\frac{\hat{G}(0)}{1+(k\xi)^2}.
\ee
This allows one to identify $\xi$ with the bulk correlation
length.
For a discussion of the structure in the presence of ions (Eq. (\ref{functional})) see Ref.
\cite{Bier2012}.

In a similar way for the case with long-ranged interaction between solvent particles (see Eq.
(\ref{functional2})) 
\begin{multline}\label{B5}
G^{-1}(\mathbf r, \mathbf r')=\delta(\mathbf r- \mathbf
r')\left(\frac{1}{\phi}+\frac{1}{1-\phi}-2\chi(T)+\frac{\beta\pi^2
A_f}{4}\right)\\-\frac{\chi(T)}{3}
\nabla^2\delta(\mathbf
r-
\mathbf r').
\end{multline}
Therefore the bulk correlation length $\xi$ for this case is given by
\be
\begin{aligned}
\frac{1}{\xi^2}
&=\frac{3}{\chi(T)}\left(\frac{1}{\phi}+\frac{1}{1-\phi}
-2\chi(T)+\frac{\pi^2\beta A_f}{4}\right),\\
& =\frac{\dps\frac{3}{\tilde\chi(T)}\Big(\frac{1}{\phi}+\frac{1}{1-\phi}-2\tilde\chi(T)\Big)}
     {\dps 1+\frac{\pi^2\beta A_f}{8\tilde\chi(T)}}
\end{aligned}
\ee
with $\tilde\chi(T)=\chi(T)-\pi^2\beta A_f/8$.

\section{First-order perturbation theory for including the long-ranged
interactions}\label{1PT}
The total grand canonical functional is given by
\be
\Omega[\phi(\mathbf r),\rho_\pm(\mathbf r)]=\Omega_0[\phi(\mathbf r),\rho_\pm(\mathbf
r)]+\Delta\Omega[\phi(\mathbf r)]
\ee
with $\Omega_0[\phi(\mathbf r),\rho_\pm(\mathbf r)]$ given by Eq.
(\ref{functional}) whereas 
$\Delta\Omega[\phi(\mathbf r)]$ is given by Eq. (\ref{DeltaOmega}) and 
depends only on $\phi(\mathbf
r)$. We choose a dimensionless coupling
parameter  $\lambda\in[0,1]$ such that for $\lambda=0$ the perturbation $\Delta\Omega$
is absent and for $\lambda=1$ the 
perturbation is fully present. The perturbed grand canonical functional is 
\be\label{PGC}
\Omega_{\lambda}[\phi(\mathbf r),\rho_\pm(\mathbf r)]=\Omega_0[\phi(\mathbf r),\rho_\pm(\mathbf
r)]+\lambda\Delta\Omega[\phi(\mathbf r)],
\ee
where $\lambda$ acts as an amplitude multiplying both $\bar w(|\mathbf r-\mathbf
r'|)$ and $V(\mathbf r)$ (see Eq. (\ref{DeltaOmega})). The equilibrium densities
$\phi_\lambda(\mathbf r)$ and
$\rho_{\pm,\lambda}(\mathbf r)$ minimize $\Omega_{\lambda}$:
\be
\frac{\delta\Omega_{\lambda}}{\delta
\phi(\mathbf
r)}[\phi_\lambda(\mathbf r),\rho_{\pm,\lambda}(\mathbf r)]=0,\ \
\frac{\delta\Omega_\lambda}{\delta \rho_{\pm}(\mathbf
r)}[\phi_\lambda(\mathbf
r),\rho_{\pm,\lambda}(\mathbf r)]=0.
\ee
Furthermore, the equilibrium densities
$\phi^{(0)}(\mathbf r)\equiv\phi_{\lambda=0}(\mathbf{r})$ and
$\rho^{(0)}_\pm(\mathbf r)\equiv\rho_{\pm,\lambda=0}(\mathbf{r})$ which
minimize $\Omega_0[\phi(\mathbf
r),\rho_\pm(\mathbf r)]$ are known (see Sec. \ref{sr}).

In order to proceed we write the equilibrium densities $\phi_\lambda(\mathbf
r)$ and $\rho_{\pm,\lambda}(\mathbf r)$ as power series in terms of $\lambda$,
\be\label{c4}
\begin{aligned}
 \phi_\lambda(\mathbf r)&=\sum_{n=0}^\infty\lambda^n\phi^{(n)}(\mathbf
r)=\phi^{(0)}(\mathbf
r)+\mathcal{O}(\lambda),\\
\rho_{\pm,\lambda}(\mathbf
r)&=\sum_{n=0}^\infty\lambda^n\rho_{\pm}^{(n)}(\mathbf
r)=\rho_\pm^{(0)}(\mathbf
r)+\mathcal{O}(\lambda),
\end{aligned}
\ee
and perform a functional Taylor expansion of the grand canonical potential 
$\Omega_{\lambda}[\phi_\lambda(\mathbf
r),\rho_{\pm,\lambda}(\mathbf r)]$ around
$\phi^{(0)}(\mathbf r)$, $\rho_\pm^{(0)}(\mathbf r)$:
\begin{widetext}
 \be\label{ome_TE}
\begin{aligned}
\Omega_{\lambda}[\phi_\lambda(\mathbf r),\rho_{\pm,\lambda}(\mathbf
r)]&=\Omega_0\left[\phi^{(0)}(\mathbf
r)+\left(\phi_\lambda(\mathbf r)-\phi^{(0)}(\mathbf
r)\right),\rho^{(0)}_\pm(\mathbf
r)+\left(\rho_{\pm,\lambda}(\mathbf r)-\rho^{(0)}_\pm(\mathbf
r)\right)\right]\\
&+\lambda\Delta\Omega\left[\phi^{(0)}(\mathbf r)+\left(\phi_\lambda(\mathbf r)-\phi^{(0)}(\mathbf
r)\right)\right]\\
&=\Omega_0[\phi^{(0)}(\mathbf
r),\rho^{(0)}_\pm(\mathbf r)]+\lambda\Delta\Omega[\phi^{(0)}(\mathbf r)]\\&+\int
d^3r\left\{\frac{\delta\Omega_{0}[\phi(\mathbf r),\rho_\pm(\mathbf r)]}{\delta\phi(\mathbf
r)}\biggr|_{\phi^{(0)}(\mathbf r),\rho_\pm^{(0)}(\mathbf
r)}(\phi_\lambda(\mathbf r)-\phi^{(0)}(\mathbf
r))\right.\\
&+\frac{\delta\Omega_{0}[\phi(\mathbf r),\rho_\pm(\mathbf r)]}{\delta\rho_+(\mathbf
r)}\biggr|_{\phi^{(0)}(\mathbf r),\rho_\pm^{(0)}(\mathbf r)}(\rho_{+,\lambda}(\mathbf
r)-\rho_+^{(0)}(\mathbf
r))\\
&\left.+\frac{\delta\Omega_{0}[\phi(\mathbf r),\rho_\pm(\mathbf
r)]}{\delta\rho_-(\mathbf
r)}\biggr|_{\phi^{(0)}(\mathbf r),\rho_\pm^{(0)}(\mathbf r)}(\rho_{-,\lambda}(\mathbf
r)-\rho_-^{(0)}(\mathbf
r))\right\}+\mathcal{O}(\lambda^2)\\
&=\Omega_0[\phi^{(0)}(\mathbf
r),\rho^{(0)}_\pm(\mathbf r)]+\lambda\Delta\Omega[\phi^{(0)}(\mathbf r)]+\mathcal{O}(\lambda^2)\\
&=\Omega_{\lambda}[\phi^{(0)}(\mathbf r),\rho^{(0)}_\pm(\mathbf r)]+\mathcal{O}(\lambda^2)
\end{aligned}
\ee
\end{widetext}
with $\phi_\lambda(\mathbf r)-\phi^{(0)}(\mathbf
r)=\mathcal{O}(\lambda)$ and $\rho_{\pm,\lambda}(\mathbf r)-\rho^{(0)}_\pm(\mathbf
r)=\mathcal{O}(\lambda)$ (see Eq. (\ref{c4})). In Eq. (\ref{ome_TE}) the first derivatives vanish
because $\phi^{(0)}$ and
$\rho_\pm^{(0)}$ minimize $\Omega_0$.
Hence $\Omega_{\lambda=1}\left[\phi_{\lambda=1}(\mathbf r),\rho_{\pm,\lambda=1}(\mathbf
r)\right]\approx\Omega_{\lambda=1}[\phi^{(0)}(\mathbf
r),\rho^{(0)}_\pm(\mathbf r)]=\Omega[\phi^{(0)}(\mathbf
r),\rho^{(0)}_\pm(\mathbf r)]$ up to second order in $\lambda$.

\section{Coefficients for the effective interface potential in the presence of
long-ranged
interactions}\label{clr}
The effective interface potential for the model with long-ranged interaction
(Sec. \ref{lr}) is
calculated by following the procedure described in Sec. \ref{sr}. The double
integrals in Eq. (\ref{DeltaOmega}) have been evaluated by performing an
asymptotic
expansion for $\ell\to\infty$.
The analytic expressions for the coefficients in Eq. (\ref{omega2}) are given by (see Appendix
\ref{integrals})

 \be
a_1(T)=\frac{(\phi_l-\phi_g)}{2}\beta(u_3\rho_w-t_3\phi_l)\label{D1}
\ee
and
\be
\begin{aligned}
 b_1(T)&=\frac{\beta(\phi_l-\phi_g)}{3}\left(\vphantom{frac{ \beta(h_1-g\phi_l)}{\beta
g+\frac{\chi(T)}{3\xi}}}\rho_wu_4-3t_3\phi_ld_w\right.\\
&\left.+3\xi
t_3\exp\left(-\frac{d_w}{\xi}\right)\left(\frac{ \beta(h_1-g\phi_l)}{\beta
g+\frac{\chi(T)}{3\xi}}
\right)\right).\label{D2}
\end{aligned}
\ee

The coefficients $a_1$ and $b_1$ can be compared with the general expressions obtained in Ref.
\cite{SDMN} within a systematic study of wetting transitions of a simple one-component fluid,
inter alia including the
presence of van der Waals tails. There the effective interface potential is expressed in terms of
the interfacial profiles which
emerge as a consequence of wetting phenomena, i.e., the wall-liquid and the free liquid-gas
interface for wetting of the wall-gas interface. Within that approach the
effective interface potential at coexistence is given by
\be
\omega(\ell)=\sum_{k=2}^{4}\bar a_k\ell^{-k}+\mathcal{O}(\ell^{-5}\ln\ell)
\ee
where
\begin{align}
 \bar a_2&=\frac{1}{2}(\phi_l-\phi_g)(\rho_wu_3-\phi_lt_3),\label{a2SD} \\ 
\bar a_3&=\bar a_3^{(0)}-2\bar a_2d_{lg}^{(1)},\label{a3SD}
\end{align}
and
\be
\bar a_4=\bar a_4^{(0)}-3a_3d_{lg}^{(1)}+3\bar a_2[d_{lg}^{(2)}-2(d_{lg}^{(1)})^2],
\ee
with
\be
\bar a_3^{(0)}=\frac13(\phi_l-\phi_g))[\rho_wu_4-\phi_l(t_4+3t_3d_{wl}^{(1)})]
\ee
and
\be
\bar a_4^{(0)}=\frac14(\phi_l-\phi_g))[\rho_wu_5-\phi_l(t_5+4t_4d_{wl}^{(1)}+6t_3d_{wl}^{(2)})].
\ee
The coefficients $t_3$, $t_4$ for the present model are given by Eqs. (\ref{t3}) and (\ref{t4});
$d_{wl}^{(i)}$ and $d_{lg}^{(i)}$  are moments of the wall-liquid interface profile $\phi_{wl}(z)$
and of the free liquid-gas interface profile $\phi_{lg}(z)$, respectively:
\be\label{dwl1}
d_{wl}^{(i)}=i\int^\infty_0 dz z^{i-1}\left[1-\frac{\phi_{wl}(z)}{\phi_l}\right],\ \ \ i=1,2,
\ee 
and
\be\label{dlg1}
d_{lg}^{(i)}=\frac{i}{\phi_l-\phi_g}\int^\infty_{-\infty} dz
z^{i-1}\left[\phi_{lg}(z)-\phi_{lg}^{sk}(z)\right],\ \ \ i=1,2.
\ee
The wall-liquid and the free liquid-gas interface profile can be calculated within our approach by
following a procedure analogous to the one described in Sec. \ref{sr}. To this end, in the case of
the
wall-liquid interface for the pure solvent, the Taylor expansion up to second order of the local
part of
the
functional in Eq. (\ref{functional}) with $\rho_\pm=0$ and $\mathbf D=0$, is performed about
$\phi_{wl}(z)=\phi_l\Theta(z-d_w)$ for $z\ge0$ where $\Theta(x)$ is the Heaviside
function. This
leads to the wall-liquid density profile
\be
\phi_{wl}(z)=\left(\phi_l+\frac{\beta(h_1-g\phi_l)}{\beta
g+\frac{\chi(T)}{3\xi}}\exp(-z/\xi)\right)\Theta(z-d_w).
\ee
Within the approximation discussed in Appendix \ref{1PT}, this expression, obtained from
minimizing $\Omega_0$ and shifting by $d_w$, is inserting into the general expression in Eq.
(\ref{dwl1}) corresponding to $\Delta\Omega$ and yields
\be\label{dwl}
d_{wl}^{(1)}=d_w-\frac{\beta\xi(h_1-g\phi_l)}{\phi_l\left(\beta
g+\frac{\chi(T)}{3\xi}\right)}\exp(-d_w/\xi).
\ee
As expected, Eq. (\ref{dwl}) respects the expected property that in the sharp-kink limit
(i.e, vanishing interfacial width $\xi$) $d_{wl}^{(1)}$ reduces to $d_w$. With this result Eq.
(\ref{D2}) can be rewritten as
$b_1(T)=\left(\beta/3\right)(\phi_l-\phi_g)\left(\rho_wu_4-3t_3\phi_ld_{wl}^{(1)}\right)$.
For the free
liquid-gas interface, we consider the functional in Eq. (\ref{solvent}). All integrals
extend over a macroscopic volume. We impose the boundary conditions
$\phi_{lg}(z\to-\infty)=\phi_l$
and $\phi_{lg}(z\to\infty)=\phi_g$. Accordingly, the Taylor expansion up to second order of the
local part of the functional is
performed
about the \textit{s}harp-\textit{k}ink profile
\be
\phi_{lg}^{sk}(z)=
\begin{cases}
\phi_l, & z<0,\\
\phi_g, & z>0.
\end{cases}
\ee
The resulting liquid-gas density profile based on $\Omega_0$ is
\be
\phi_{lg}(z)=
\begin{cases}
\phi_l+\frac{\phi_g-\phi_l}{2}\exp(z/\xi), & z<0,\\
\phi_g+\frac{\phi_l-\phi_g}{2}\exp(-z/\xi), & z>0.
\end{cases}
\ee
Again, within the approximation discussed in Appendix \ref{1PT}, this profile stemming from
$\Omega_0$ is inserted into the general expression in Eq. (\ref{dlg1}), which is based on
$\Delta\Omega$, and renders
\be\label{dlg}
d_{lg}^{(1)}=0.
\ee
Inserting Eqs. (\ref{t3}), (\ref{t4}), (\ref{dwl}), and (\ref{dlg}) into Eqs. (\ref{a2SD})
and (\ref{a3SD}) one obtains (with $t_4=0$)
\be
\beta\bar a_2=a_1\ \  \text{and}\ \ \beta\bar a_3=b_1.
\ee
This leads to the satisfactory statement that if the general results in Ref. \cite{SDMN} for the
effective interface potential are applied to the present model one finds the same effective
interface potential as the one obtained directly within the present model.
\begin{widetext}
\section{Derivation of the effective interface potential for the model with
long-ranged interactions}\label{integrals}

The derivation of  $\omega(\ell)$ in Eq. (\ref{omega2}) follows the same procedure as described in
Sec
II. We perform a Taylor expansion up to second order of the local part of the functional in Eq.
(\ref{functional2}) 
about the sharp-kink profile in Eq. (\ref{sharpkinkphi}) shifted by $d_w$ and the
sharp-kink profile in Eq. (\ref{sharpkinkI})  with the bulk
state being determined by Eq. (\ref{bulk_lr}) below. From this
expansion
we obtain an approximate variational functional $\hat \Omega_{lr}$ for the model with
\textit{l}ong-\textit{r}anged interactions. By subtracting the bulk contribution
$\Omega_{b,lr}$ of the gas phase we obtain the surface contribution $\Omega_{s,lr}$ to
this variational
functional:
\be\label{func_approx-lg}
\begin{aligned}
\beta\Omega_{s,lr}\left(\ell,[\Delta\phi(z),\Delta\rho_\pm(z)]\right)&=\frac{
\beta\left(\hat\Omega_{lr}(\ell , [ \Delta\phi(z) ,
\Delta\rho_\pm(z)])-V\Omega_{b,lr}(\phi_g,0)\right)}{A}\\
&=\beta\ell\left[\Omega_{b,lr}(\phi_l,I)-\Omega_{b,lr}(\phi_g,0)\right]-\beta
d_w\Omega_{b,lr}(\phi_l,I)\\
&+\int_{d_w}^\ell\!\!
dz\Biggl\{\frac{\chi(T)}{6}{\left(\!\frac{d}{dz}
\Delta\phi(z)\!\right)\!}^2\!\!+\!\frac12\left(\Delta\phi(z)\right)^2\left(\frac
{1}{\phi_l}\!+\!\frac {1}{1\!-\!\phi_l}\!-\!2\chi(T)\right)\!\Biggr\}\\
&+\int_\ell^\infty\!\!\!
dz\Biggl\{\frac{\chi(T)}{6}
{\left(\!\frac{d}{dz}\Delta\phi(z)\!\right)\!}
^2\!\!+\!\frac12\left(\Delta\phi(z)\right)^2\left(\frac { 1 } { \phi_g
}\!+\!\frac{1}{1\!-\!\phi_g}\!-\!2\chi(T)\right)\!\Biggr\}\\
& -\beta
h_1\phi_l-\beta
h_1\Delta\phi(d_w)+\beta\frac{g}{2}(\phi_l+\Delta\phi(d_w))^2\\
&-\frac12\phi_l^2\beta\left(\mathrm{I_0}_{(d_w,-\infty)}^{(\ell,d_w)}+\mathrm{I_0}_{(d_w,\ell)}^{
(\ell ,
\infty)}\right)+\phi_l\phi_g\beta\mathrm{I_0}_{(d_w,\ell)}^{(\ell,\infty)}
-\frac12\phi_g^2\beta\mathrm { I_0 } _ {
(\ell,-\infty)}^{(\infty,\ell)}\\
&-\phi_l\beta\mathrm{I_2}_{(d_w,-\infty)}^{(\ell,
d_w)}-(\phi_l-\phi_g)\beta\left(\mathrm{I_2}_{(d_w,\ell)}^{(\ell,\infty)}
-\mathrm{I_2}_{(\ell,d_w)}^{(\infty,\ell)}\right)
-\phi_g\beta\mathrm{I_2}_{(\ell,-\infty)}^{(\infty,d_w)}\\
&+\frac12\beta\left(\mathrm{I_3}_{(d_w,d_w)}^{(\ell,\ell)}+\mathrm{I_3}_{(d_w,\ell)}^{(\ell,\infty)}
+\mathrm{I_3} _
{(\ell,d_w)}^{(\infty,\ell)}+\mathrm{I_3}_{(\ell,\ell)}^{(\infty,\infty)}\right)\\
&+\phi_l\rho_w\beta\int_{d_w}^\ell\!\!dzV(z)+\rho_w\beta\mathrm{I_1}_{
(d_w)}^{(\ell)}\\
&+\phi_g\rho_w\beta\int_\ell^\infty\!\!\!dzV(z)+\rho_w\beta\mathrm{I_1}_{(\ell)}^{(\infty)}\\
&+\int_0^\ell\!\!dz\Biggl\{\frac{1}{2I}\sum_{i=\pm}\left(\Delta\rho_i(z)\right)^2+\frac{2\pi
l_B}{\varepsilon_l}\left(D(z,[\Delta\rho_\pm])\right)^2\Biggr\},
\end{aligned}
\ee
where $d_w$ describes the excluded volume due to the repulsive part of the substrate potential
$V(z)$ given by Eq. (\ref{Vz}) and
\be\label{wzz'}
w(|z-z'|)=\frac{\pi A_f}{2\left[(z-z')^2+1\right]^2}.
\ee
The bulk grand canonical potential density $\Omega_{b,lr}$ per $k_BT$  is given by
\be\label{bulk_lr}
\begin{aligned}
\beta\Omega_{b,lr}(\phi,\rho)&=\phi(\ln(\phi)-\beta\mu_{\phi}
)+(1-\phi)\ln(1-\phi)+\chi(T)\phi(1-\phi)+\frac12\phi^2\int_{-\infty}^{\infty}
\!dx w(|x|)\\
&+2\rho(\ln(\rho)-1)\!-\!\beta\mu_I\rho\!+\!\rho\left(V_+(\phi)\!+\!V_-(\phi)\right).
\end{aligned}
\ee
$\mathrm{I_0}$, $\mathrm{I_1}$, $\mathrm{I_2}$, and $\mathrm{I_3}$ are abbreviations for the
following types of integrals:
\be
\mathrm{I_0}_{(u_1,v_1)}^{(u_2,v_2)}=\int_{u_1}^{u_2}dz\int_{v_1}^{v_2}dz'w(|z-z'|),
\ee
\be\label{un}
\mathrm{I_1}_{(u_1)}^{(u_2)}=\int_{u_1}^{u_2}dzV(z)\Delta\phi(z),
\ee
\be
\mathrm{I_2}_{(u_1,v_1)}^{(u_2,v_2)}=\int_{u_1}^{u_2}dz\int_{v_1}^{v_2}dz'\Delta\phi(z)w(|z-z'|),
\ee
and
\be\label{tres}
\mathrm{I_3}_{(u_1,v_1)}^{(u_2,v_2)}=\int_{u_1}^{u_2}dz\int_{v_1}^{v_2}
dz'\Delta\phi(z)\Delta\phi(z')w(|z-z'|).
\ee

The integrals in Eqs. (\ref{un})-(\ref{tres}) are evaluated at two-phase coexistence using the
solutions for $\Delta\phi(z)$ obtained in Sec. \ref{sr} (see Eqs. (\ref{solutionphi}) and
(\ref{consts_co})).
We are interested in the asymptotic behavior of these integrals in the limit $\ell\to\infty$. Using
Eqs. (\ref{solutionphi}) and (\ref{Vz}), $\mathrm{I_1}_{(d_w)}^{(\ell)}$ can be written as
\be\label{I1a}
\begin{aligned}
\mathrm{I_1}_{(d_w)}^{(\ell)}&=-\int_{d_w}^{\ell}\!\!dz\left[\sum_{i\ge3}\frac{u_i}{z^i}\right]
\!\!\left [
\vphantom{\sum_{i\ge3}\frac{u_i}{z^i}} A_ { l}\exp(z/\xi)\!+\!B_{l}\exp(-z/\xi)\right]\\
&=-A_l\int_{d_w}^{\ell}\!\!dz\left[\sum_{i\ge3}\frac{u_i}{z^i}\right]\exp(z/\xi)-B_{l}\int_{d_w}^{
\ell }\!\!
dz\left[\sum_{i\ge3}\frac{u_i}{z^i}\right]\exp(-z/\xi).
\end{aligned}
\ee

Asymptotic approximations for the integrals in Eq. (\ref{I1a}) are obtained via integrating by
parts repeatedly:
\be
\begin{aligned}
\int_{d_w}^\ell dz\frac{\exp(z/\xi)}{z^3}&=\frac{\xi\exp(\ell/\xi)}{\ell^3}-\frac{\xi\exp(d_w/\xi)}{
d_w^3}\\
&\phantom{=}+3\xi \int_{d_w}^\ell dz\frac{\exp(z/\xi)}{z^4},\\
&=\frac{\xi\exp(\ell/\xi)}{\ell^3}\!+\!\mathcal{O}(\ell^{-4}\exp(\ell/\xi)),\ \ell\!\gg\!d_w,
\end{aligned}
\ee
and
\be
\int_{d_w}^\ell dz\frac{\exp(-z/\xi)}{z^3}=\int_{d_w}^\infty
dz\frac{\exp(-z/\xi)}{z^3}-\int_{\ell}^\infty dz\frac{\exp(-z/\xi)}{z^3}
\ee
with
\be
\begin{aligned}
\int_{\ell}^\infty\!\!\!
dz\frac{\exp(-z/\xi)}{z^3}&=\frac{\xi\exp(-\ell/\xi)}{\ell^3}-3\xi \int_{\ell}^\infty
dz\frac{\exp(-z/\xi)}{z^4},\\
&=\frac{\xi\exp(-\ell/\xi)}{\ell^3}\!+\!\mathcal{O}(\ell^{-4}\exp(\ell/\xi)),\\
&\phantom{=}\ \ell\!\gg\!d_w.
\end{aligned}
\ee
Here and in the following we have used the properties
\[\int_a^{\ell+b} dz\frac{P_1(z)}{P_2(z)}\exp(z/\xi)\stackrel{\ell\to
\infty}{\simeq}\frac{c_1}{c_2}\xi\ell^{n_1-n_2}\exp(\ell/\xi)\] and \[\int_{\ell+a}^\infty
dz\frac{P_1(z)}{P_2(z)}\exp(-z/\xi)\stackrel{\ell\to
\infty}{\simeq}\frac{c_1}{c_2}\xi\ell^{n_1-n_2}\exp(-\ell/\xi)\] for two polynomials $P_1(z)$ and
$P_2(z)$ of degrees
$n_1$ and $n_2$ with the leading coefficients $c_1$ and $c_2$, respectively, of the
leading terms which follows from
L'H\^opital's rule.

For convenience, we write the expressions in Eq. (\ref{consts_co}) as
\be\label{cc}
\begin{aligned}
A_{l}&=A_1\exp(-\ell/\xi),\\
B_{l}&=B_1\exp(-\ell/\xi)+B_2,\\
B_{g}&=-A_1\exp(\ell/\xi)+B_1\exp(-\ell/\xi)+B_2,
\end{aligned}
\ee
with
\be
\begin{aligned}
A_1&=\frac{\phi_g-\phi_l}{2},\\
B_1&=\frac{\left(\frac{\chi(T)}{3\xi}-\beta g\right)\left(\phi_g-\phi_l\right)}
{2\left(\beta g+\frac{\chi(T)}{3\xi}\right)},\\
B_2&=\frac{\beta(h_1-g\phi_l)}{\beta g+\frac{\chi(T)}{3\xi}}.
\end{aligned}
\ee
Collecting only algebraic terms up to the order
$1/\ell^3$, one obtains for $\mathrm{I_1}_{(d_w)}^{(\ell)}$
\be\label{I11r}
\mathrm{I_1}_{(d_w)}^{(\ell)}=-\frac{\xi A_1u_3}{\ell^3}+\mathcal{O}\left(\frac{1}{\ell^4}
\right).
\ee

Analogously, for $\mathrm{I_1}_{(\ell)}^{(\infty)}$ one has
\be\label{I12r}
\begin{aligned}
\mathrm{I_1}_{(\ell)}^{(\infty)}&=-\int_{\ell}^{\infty}dz\left[\sum_{i\ge3}\frac{u_i}{z^i}\right]
B_{g}\exp(-z/\xi)\\
&=\frac{\xi
A_1u_3}{\ell^3}+\mathcal{O}\left(\frac{1}{\ell^4}\right).
\end{aligned}
\ee

In order to calculate integrals of the type $\mathrm{I_2}$ we first integrate  Eq. (\ref{wzz'})
using the various
integration limits appearing in Eq. (\ref{func_approx-lg}) so that
\be\label{intw0}
\int_{-\infty}^{d_w}dz'w(|z-z'|)=\frac{\pi
A_f}{4}\left[\frac{\pi}{2}-\arctan(z-d_w)-\frac{z-d_w}{(z-d_w)^2+1}\right],
 \ee
\be\label{intw1}
 \int_{\ell}^{\infty}dz'w(|z-z'|)=\frac{\pi
A_f}{4}\left[\arctan(z-\ell)+\frac{z-\ell}{(z-\ell)^2+1}+\frac{\pi}{2}\right],
 \ee
and
 \be\label{intw2}
\int_{d_w}^{\ell}dz'w(|z-z'|)=\frac{\pi
A_f}{4}\left[\arctan(z-d_w)+\frac{z-d_w}{(z-d_w)^2+1}-\arctan(z-\ell)-\frac{z-\ell}{(z-\ell)^2+1}
\right ] .
\ee

Using Eqs. (\ref{solutionphi}) and (\ref{intw0}) we can
write $\mathrm{I_2}_{(d_w,-\infty)}^{(\ell,d_w)}$ as
\be\label{I2a}
\begin{aligned}
\mathrm{I_2}_{(d_w,-\infty)}^{(\ell,d_w)}&=\frac{\pi
A_f}{4}\int_{d_w}^{\ell}\!\!\!dz\biggl[\frac{\pi}{2}-\arctan(z-d_w)-\frac{z-d_w}{(z-d_w)^2+1}\biggr]
\biggl [
A_
{l}\exp(z/\xi)+B_{l}\exp(-z/\xi)\biggr]\\
&=\frac{\pi
A_f}{4}A_{l}\left[\frac{\xi\pi}{2}
\left(\exp(\ell/\xi)-\exp(d_w/\xi)\right)-\xi\exp(\ell/\xi)\arctan(\ell-d_w)
\right.\\
&\left. +\xi\exp(d_w/\xi )\int_{0}^{\ell-d_w}
dy\frac{\exp(y/\xi)}{y^2+1}-\exp(d_w/\xi)\int_0^{\ell-d_w}
dz\frac{y\exp(y/\xi)}{y^2+1}\right]\\
&+\frac{\pi
A_f}{4}B_{l}\left[\frac{-\xi\pi}{2}
\left(\exp(-\ell/\xi)-\exp(-d_w/\xi)\right)+\xi\exp(-\ell/\xi)\arctan(\ell-d_w)\right.\\
&\left.-\xi\exp(-d_w/\xi)\int_0^{\ell-d_w}
dy\frac{\exp(-y/\xi)}{y^2+1}-\exp(-d_w/\xi)\int_0^{\ell-d_w}
dy\frac{y\exp(-y/\xi)}{y^2+1}\right
],
\end{aligned}
\ee
where we have changed the integration variable to $y=z-d_w$. Asymptotic approximations for the
integrals in Eq. (\ref{I2a}) are obtained via integrating by parts
repeatedly:
\be\label{int1}
\begin{aligned}
\int_0^{\ell-d_w}
dz\frac{\exp(z/\xi)}{z^2+1}&=\frac{\xi\exp((\ell-d_w)/\xi)}{(\ell-d_w)^2+1}-\xi+\frac{
2(\ell-d_w)\xi^2\exp((\ell-d_w)/\xi) } {
((\ell-d_w)^2+1)^2}\\
&+\xi^2\int_0^{\ell-d_w} dz\frac{6z^2-2}{(z^2+1)^3}\exp(z/\xi),\\
&=\frac{\xi\exp((\ell-d_w)/\xi)}{\ell^2}+\frac{2(d_w\xi+\xi^2)\exp
((\ell-d_w)/\xi)}{\ell^3}\\
&+\mathcal{O}(\ell^{-4}\exp(\ell/\xi)),\ \ \ \ell\gg d_w,
\end{aligned}
\ee

\be\label{int2}
\begin{aligned}
\int_0^{\ell-d_w}
dz\frac{z\exp(z/\xi)}{z^2+1}&=\frac{\xi(\ell-d_w)\exp((\ell-d_w)/\xi)}{(\ell-d_w)^2+1}-\frac{
\xi^2\exp((\ell-d_w)/\xi) } {
(\ell-d_w)^2+1}+\xi^2\\
&+\frac{2((\ell-d_w)\xi)^2\exp((\ell-d_w)/\xi)}{((\ell-d_w)^2+1)^2}-\frac{
6(\ell-d_w)\xi^3\exp((\ell-d_w)/\xi)
}{((\ell-d_w)^2+1)^2}\\
&+\frac{8((\ell-d_w)\xi)^3\exp((\ell-d_w)/\xi)}{((\ell-d_w)^2+1)^3}+\int_0^{\ell-d_w}\!\!\!
dz\frac{6(z^4-6z^2+1)}
{(z^2+1)^4}\xi^3\exp(z/\xi) , \\
&=\xi\exp((\ell-d_w)/\xi)\left(\frac{1}{\ell}+\frac{d_w}{\ell^2}+\frac{d_w^2-1}{\ell^3}
\right)+\frac{\xi^2\exp((\ell-d_w)/\xi)}{\ell^2}\\
&+\frac{2\xi^2(d_w+\xi)\exp((\ell-d_w)/\xi)}{\ell^3}+\mathcal{O}(\ell^{-4}\exp(\ell/\xi)),\ \
\ \ell\gg d_w,
\end{aligned}
\ee
and
\be
\int_0^{\ell-d_w}
dz\frac{\exp(-z/\xi)}{z^2+1}=\int_0^\infty
dz\frac{\exp(-z/\xi)}{z^2+1}-\int_{\ell-d_w}^\infty
dz\frac{\exp(-z/\xi)}{z^2+1}
\ee
with
\be\label{Intwithoutz}
\begin{aligned}
\int_{\ell-d_w}^\infty
dz\frac{\exp(-z/\xi)}{z^2+1}&=\frac{
\xi\exp(-(\ell-d_w)/\xi)}{(\ell-d_w)^2+1}-\frac{2(\ell-d_w)\xi^2\exp(-(\ell-d_w)/\xi)}{
((\ell-d_w)^2+1)^2 }\\
&+\int_{\ell-d_w}^\infty dz\frac{6z^2-2}{(z^2+1)^3}\xi^2\exp(-z/\xi)\\
&=\frac{\xi\exp(-(\ell-d_w)/\xi)}{\ell^2}+\frac{2\xi(d_w-\xi)\exp
(-(\ell-d_w)/\xi)}{\ell^3}\\
&+\mathcal{O}(\ell^{-4}\exp(-\ell/\xi)),\ \ \ \ell\gg d_w,
\end{aligned}
\ee
and
\be
\int_0^{\ell-d_w}
dz\frac{z\exp(-z/\xi)}{z^2+1}=\int_0^\infty
dz\frac{z\exp(-z/\xi)}{z^2+1}-\int_{\ell-d_w}^\infty
dz\frac{z\exp(-z/\xi)}{z^2+1}
\ee
with
\be\label{Intwithz}
\begin{aligned}
\int_{\ell-d_w}^\infty
dz\frac{z\exp(-z/\xi)}{z^2+1}&=\frac{\xi(\ell-d_w)\exp(-(\ell-d_w)/\xi)}{
(\ell-d_w)^2+1}+\frac{\xi^2\exp(-(\ell-d_w)/\xi)}{
\ell^2+1}\\
&-\frac{2((\ell-d_w)\xi)^2\exp(-(\ell-d_w)/\xi)}{((\ell-d_w)^2+1)^2}
-\frac{6(\ell-d_w)\xi^3\exp(-(\ell-d_w)/\xi)}{((\ell-d_w)^2+1)^2}\\
&+\frac{8((\ell\!-\!d_w)\xi)^3\exp(-(\ell\!-\!d_w)/\xi) } { ((\ell-d_w)^2+1)^3 }
-\xi^3\int_{\ell-d_w}^{\infty}\!\!\!\!
dz\frac{6(z^4\!-\!6z^2\!+\!1)}
{(z^2\!+\!1)^4}\exp(-z/\xi), \\
&=\xi\exp(-(\ell-d_w)/\xi)\left(\frac{1}{\ell}+\frac{d_w}{\ell^2}+\frac{d_w^2-1}{\ell^3}
\right)-\frac{\xi^2\exp(-(\ell-d_w)/\xi)}{ \ell^2}\\
&+\frac{2\xi^2(\xi-d_w)\exp(-(\ell-d_w)/\xi)}{\ell^3}+\mathcal{O}(\ell^{-4}\exp(-\ell/\xi)),\ \ \
\ell\gg d_w.
\end{aligned}
\ee

Additionally, for $\ell\gg d_w$ one has
\be\label{arctan}
\frac{\pi}{2}-\arctan(\ell-d_w)=\frac{1}{\ell}+\frac{d_w}{\ell^2}+\frac{d_w^2-1/3}{\ell^3}+\mathcal{
O } \left(\frac{1}{\ell^4}
\right).
\ee

Note that Eqs. (\ref{int1}) and (\ref{int2}) contain terms which increase exponentially
with $\ell$.
However, these two integrals are multiplied by $A_{l}$ which decays exponentially with $\ell$
(see Eqs. (\ref{I1a}) and (\ref{cc})).

Collecting constants and algebraic terms up to the order
$1/\ell^3$, for $\mathrm{I_2}_{(d_w,-\infty)}^{(\ell,d_w)}$ one obtains 

\be\label{I21r}
\begin{aligned}
\mathrm{I_2}_{(d_w,-\infty)}^{(\ell,d_w)}&=\frac {A_f\pi B_2}
{4}\exp(-d_w/\xi)\left(\frac{\xi\pi}{2}\!-\xi\!\int_0^\infty
\!dy\frac{\exp\left(-\frac{y}{\xi}\right)}{y^2+1}\!-\!\int_0^\infty
\!dy\frac{y\exp\left(-\frac{y}{\xi}\right)} {y^2+1}\right)\\
&+\frac{\pi A_f\xi A_1}{6\ell^3}+\mathcal{O}\left(\frac{1}{\ell^4}\right).
\end{aligned}
\ee
Similarly, $\mathrm{I_2}_{(\ell,-\infty)}^{(\infty,d_w)}$ can be written as
\be\label{I22r}
\begin{aligned}
\mathrm{I_2}_{(\ell,-\infty)}^{(\infty,d_w)}&=\frac{\pi
A_f}{4}\int_{\ell}^{\infty}\!\!\!dz\biggl[\frac{\pi}{2}-\arctan(z-d_w)-\frac{z-d_w}{(z-d_w)^2+1}
\biggr ] B_{g}\exp(-z/\xi)\\
&=-\frac{\pi
A_f\xi A_1}{6\ell^3}+\mathcal{O}\left(\frac{1}{\ell^4}\right).
\end{aligned}
\ee

Using Eqs. (\ref{solutionphi}) and (\ref{intw1}) we can
write $\mathrm{I_2}_{(d_w,\ell)}^{(\ell,\infty)}$ as
\be\label{I2b}
\begin{aligned}
\mathrm{I_2}_{(d_w,\ell)}^{(\ell,\infty)}&=\frac{\pi
A_f}{4}\int_{d_w}^{\ell}dz\biggl[\arctan(z-\ell)+\frac{z-\ell}{(z-\ell)^2+1}+\frac{\pi}{2}\biggr]
\biggl [
A_
{l}\exp(z/\xi)+B_{l}\exp(-z/\xi)\biggr]\\
&=\frac{\pi
A_f}{4}A_{l}\left[\frac{\xi\pi}{2}
\left(\exp(\ell/\xi)-\exp(d_w/\xi)\right)+\xi\exp(d_w/\xi)\arctan(\ell-d_w)
\right.\\
&\left. -\xi\exp(\ell/\xi )\int_0^{\ell-d_w}
dy\frac{\exp(-y/\xi)}{y^2+1}-\exp(\ell/\xi)\int_0^{\ell-d_w}
dz\frac{y\exp(-y/\xi)}{y^2+1}\right]\\
&+\frac{\pi
A_f}{4}B_{l}\left[\frac{-\xi\pi}{2}
\left(\exp(-\ell/\xi)-\exp(-d_w/\xi)\right)-\xi\exp(-d_w/\xi)\arctan(\ell-d_w)\right.\\
&\left.+\xi\exp(-\ell/\xi)\int_0^{\ell-d_w}
dy\frac{\exp(y/\xi)}{y^2+1}-\exp(-\ell/\xi)\int_0^{\ell-d_w}
dy\frac{y\exp(y/\xi)}{y^2+1}\right
],
\end{aligned}
\ee
where we have changed the integration variable to $y=\ell-z$. Using Eqs.
(\ref{int1})-(\ref{arctan}) one obtains asymptotically
\be\label{I2ar}
\begin{aligned}
\mathrm{I_2}_{(d_w,\ell)}^{(\ell,\infty)}&=\frac {A_f\pi A_1}
{4}\left(\frac{\xi\pi}{2}\!-\xi\!\int_0^\infty
\!dy\frac{\exp\left(-\frac{y}{\xi}\right)}{y^2+1}\!-\!\int_0^\infty
\!dy\frac{y\exp\left(-\frac{y}{\xi}\right)} {y^2+1}\right)\\
&+\frac{
A_f\xi\pi B_2} {6\ell^3}\exp(-d_w/\xi)+\mathcal{O}(\frac{1}{\ell^4}).
\end{aligned}
\ee
Similarly, $\mathrm{I_2}_{(\ell,d_w)}^{(\infty,\ell)}$ can be written as (see Eqs.
(\ref{solutionphi})
and (\ref{intw2}))
\be\label{I2c}
\begin{aligned}
\mathrm{I_2}_{(\ell,d_w)}^{(\infty,\ell)}&=\int_\ell^\infty\!dz\frac{\pi
A_f}{4}\biggl[\arctan(z-d_w)+\frac{z-d_w}{(z-d_w)^2+1}-\arctan(z-\ell)-\frac{z-\ell}{(z-\ell)^2+1}
\biggr ]
\times\\
&\times\biggl[B_ {l}-A_{l}\exp(2\ell/\xi)\biggr]\exp(-z/\xi)\\
&=\frac{\pi
A_f}{4}\left[\vphantom{\int_0^\infty dy\frac{\exp(-y/\xi)}{y^2+1}}B_
{l}-A_{l}\exp(2\ell/\xi)\right]\left[\vphantom{\int_0^\infty
dy\frac{\exp(-y/\xi)}{y^2+1}}\xi\arctan(\ell-d_w)\exp(-\ell/\xi)\right.\\
&+\xi\exp(-d_w/\xi)\int_{\ell-d_w}^\infty
dy\frac{\exp(-y/\xi)}{y^2+1}+\exp(-d_w/\xi)\int_{\ell-d_w}^\infty
dy\frac{y\exp(-y/\xi)}{y^2+1}\\
&\left.-\xi\exp(-\ell/\xi)\int_0^\infty dy'\frac{\exp(-y'/\xi)}{y'^2+1}-\exp(-\ell/\xi)\int_0^\infty
dy'\frac{y'\exp(-y'/\xi)}{y'^2+1}\right],
\end{aligned}
\ee
where we have changed the integration variable to $y=z-d_w$ and $y'=z-\ell$, respectively.
Using Eqs.
(\ref{int1})-(\ref{arctan}), this leads to the asymptotic behavior
\be\label{I2br}
\begin{aligned}
\mathrm{I_2}_{(\ell,0)}^{(\infty,\ell)}&=-\frac { A_f\pi A_1}
{4}\left(\frac{\xi\pi}{2}\!-\xi\!\int_0^\infty
\!dy\frac{\exp\left(-\frac{y}{\xi}\right)}{y^2+1}\!-\!\int_0^\infty
\!dy\frac{y\exp\left(-\frac{y}{\xi}\right)} {y^2+1}\right)\\
&+\frac{\pi
A_f\xi A_1}{6\ell^3}+\mathcal{O}(\frac{1}{\ell^4}).
\end{aligned}
\ee

Integrals of the type  $\mathrm{I_3}$ can be written as (see Eqs.
(\ref{solutionphi}) and Eqs. (\ref{wzz'}))
\be\label{I3a}
\begin{aligned}
\mathrm{I_3}_{(d_w,d_w)}^{(\ell,\ell)}&=\frac
{A_f\pi}{2}\left\{A_{l}^2\int_{d_w}^\ell\!dz\int_{d_w}^\ell\!dz'\frac{\exp(z/\xi)\exp(z'/\xi)}{\left
[
(z-z')^2+1\right] ^2}\right.\\
&+2A_{l}B_{l}\int_{d_w}^\ell\!dz\int_{d_w}^\ell\!dz'\frac{\exp(z/\xi)\exp(-z'/\xi)}{\left[
(z-z')^2+1\right]^2}\\
&\left.+B_{l}^2\int_{d_w}^\ell\!dz\int_{d_w}^\ell\!dz'\frac{\exp(-z/\xi)\exp(-z'/\xi)}{\left[
(z-z')^2+1\right ] ^2}\right\}, 
\end{aligned}
\ee
\be\label{I3b}
\begin{aligned}
\mathrm{I_3}_{(\ell,\ell)}^{(\infty,\infty)}&=\frac
{A_f\pi}{2}B_{g}^2\int_\ell^\infty\!dz\int_\ell^\infty\!dz'\frac{\exp(-z/\xi)\exp(-z'/\xi)}{
\left [
(z-z')^2+1\right] ^2},
\end{aligned}
\ee

\be\label{I3c}
\begin{aligned}
\mathrm{I_3}_{(d_w,\ell)}^{(\ell,\infty)}&=\frac
{A_f\pi}{2}\left\{A_{l}B_{g}\int_{d_w}^\ell\!dz\int_\ell^\infty\!dz'\frac{\exp(z/\xi)\exp(-z'/\xi)
} { \left [
(z-z')^2+1\right] ^2}\right.\\
&\left.+B_{l}B_{g}\int_{d_w}^\ell\!dz\int_\ell^\infty\!dz'\frac{\exp(-z/\xi)\exp(-z'/\xi)}{\left[
(z-z')^2+1\right ] ^2}\right\}, 
\end{aligned}
\ee
and 
\be\label{I3c_1}
\begin{aligned}
\mathrm{I_3}_{(\ell,d_w)}^{(\infty,\ell)}&=\frac
{A_f\pi}{2}\left\{A_{l}B_{g}\int_{\ell}^\infty\!dz\int_{d_w}^\ell\!dz'\frac{
\exp(-z/\xi)\exp(z'/\xi)
} { \left [
(z-z')^2+1\right] ^2}\right.\\
&\left.+B_{l}B_{g}\int_{\ell}^\infty\!dz\int_{d_w}^\ell\!dz'\frac{\exp(-z/\xi)\exp(-z'/\xi)}{
\left [
(z-z')^2+1\right ] ^2}\right\}, 
\end{aligned}
\ee
with $\mathrm{I_3}_{(d_w,\ell)}^{(\ell,\infty)}=\mathrm{I_3}_{(\ell,d_w)}^{(\infty,\ell)}$.
The double integrals can be reduced to single ones as follows:
\be
\begin{aligned}
\int_{u_1}^{u_2}\!dz\int_{v_1}^{v_2}\!dz'\frac{\exp(z/\xi)\exp(z'/\xi)}{\left[
(z-z')^2+1\right]^2}
&=\int_{u_1}^{u_2}\!dz\exp(2z/\xi)\int_{v_1}^{v_2}\!dz'\frac{\exp((z'-z)/\xi)}{
\left[(z-z')^2+1\right] ^2}\\
&\overset{(y:=z'\!-\!z)}{=}\int_{u_1}^{u_2}\!dz\exp(2z/\xi)\int_{v_1-z}^{v_2-z}\!dy\frac{
\exp(y/\xi)}{\left[
y^2+1\right] ^2}\\
&=\left.\frac{\xi}{2}\left[\exp(2z/\xi)\int_{v_1-z}^{v_2-z}\!dy\frac{
\exp(y/\xi)}{\left[y^2+1\right]^2}\right]\right|_{z=u_1}^{z=u_2}\\
&-\frac{\xi}{2}\int_{u_1}^{u_2}\!dz\exp(2z/\xi)\left[
-\frac{\exp((v_2-z)/\xi)}{\left[(v_2-z)^2+1\right]^2}\right.\\
&+\left.\frac{\exp((v_1-z)/\xi)}{\left[
(v_1-z)^2+1\right]^2}\right] \\
&{(y:=z\!-\!v_{2,1})}{=}\left.\frac{\xi}{2}\left[\exp(2z/\xi)\int_{v_1-z}^{v_2-z
} \!dy\frac
{
\exp(y/\xi)}{\left[y^2+1\right]^2}\right]\right|_{z=u_1}^{z=u_2}\\
&\qquad\qquad+\left.\frac{\xi}{2}\left[\exp(2z/\xi)\int_{
u_1-z}^{u_2-z}\!dy\frac{
\exp(y/\xi)}{\left(y^2+1\right)
^2}\right]\right|_{z=v_1}^{z=v_2},
\end{aligned}
\ee
\be
\begin{aligned}
\int_{u_1}^{u_2}\!dz\int_{v_1}^{v_2}\!dz'\frac{\exp(-z/\xi)\exp(-z'/\xi)}{\left[
(z-z')^2+1\right]^2}
&=-\left.\frac{\xi}{2}\left[\exp(-2z/\xi)\int_{v_1-z}^{v_2-z}\!dy\frac
{\exp(-y/\xi)}{\left(y^2+1\right)^2}\right]\right|_{z=u_1}^{z=u_2}\\
&-\left.\frac{\xi}{2}\left[\exp(-2z/\xi)\int_{
u_1-z}^{u_2-z}\!dy\frac{
\exp(-y/\xi)}{\left(
y^2+1\right)
^2}\right]\right|_{z=v_1}^{z=v_2},
\end{aligned}
\ee
and
\be
\begin{aligned}
\int_{u_1}^{u_2}\!dz\int_{v_1}^{v_2}\!dz'\frac{\exp(z/\xi)\exp(-z'/\xi)}{\left[
(z-z')^2+1\right]
^2}&=\left[\left.z\int_{v_1-z}^{v_2-z}\!dy\frac
{\exp(-y/\xi)}{\left[y^2+1\right]^2}\right]\right|_{z=u_1}^{z=u_2}\\
&\left.+\left[\int_{u_1-z}^{u_2-z}\!dy(y+z)\frac{
\exp(y/\xi)}{\left[
y^2+1\right]
^2}\right]\right|_{z=v_1}^{z=v_2}\\
&=\left[\left.z\int_{v_1-z}^{v_2-z}\!dy\frac
{\exp(-y/\xi)}{\left[y^2+1\right]^2}\right]\right|_{z=u_1}^{z=u_2}\\
&+\left[\left.z\int_{u_1-z}^{
u_2-z }
\!dy\frac
{\exp(y/\xi)}{\left[y^2+1\right]^2}\right]\right|_{z=v_1}^{z=v_2}\\
&\left.+\left[\int_{u_1-z}^{u_2-z}\!dy\frac{y
\exp(y/\xi)}{\left[
y^2+1\right]
^2}\right]\right|_{z=v_1}^{z=v_2}.
\end{aligned}
\ee
Inserting these expressions into Eqs. (\ref{I3a})-(\ref{I3c}) leads to
\be\label{I3a1}
\begin{aligned}
\mathrm{I_3}_{(d_w,d_w)}^{(\ell,\ell)}&=\frac
{A_f\pi}{2}\left\{A_{l}^2\xi\left[\exp(2\ell/\xi)\int_0^{\ell-d_w}\!dy\frac{
\exp(-y/\xi)} { \left(
y^2+1\right)^2}-\exp(2d_w/\xi)\int_{0}^{\ell-d_w}\!dy\frac{
\exp(y/\xi)} { \left(y^2+1\right)^2}\right]\right.\\
&+2A_{l}B_{l}\left[(\ell-d_w)\left(\int_{0}^{\ell-d_w}\!dy\frac{\exp(-y/\xi)}{\left(y^2+1\right)^2} 
+\int_{0}^{\ell-d_w}\!dy\frac{\exp(y/\xi)}{\left(y^2+1\right)^2}\right)
\right.\\
&-\int_{0}^{\ell-d_w}\!dy\frac{y\exp(-y/\xi)}{\left(y^2+1\right)^2}\left.-\int_{0}^{
\ell-d_w}\!dy\frac {
y\exp(y/\xi)}{\left(y^2+1\right)^2}\right]\\
&\left.+B_{l}^2\xi\left[-\exp(-2\ell/\xi)\int_{0}^{\ell-d_w}\!dy\frac{
\exp(y/\xi)} { \left(
y^2+1\right)^2}+\exp(-2d_w/\xi)\int_{0}^{\ell-d_w}\!dy\frac{
\exp(-y/\xi)} { \left(y^2+1\right)^2}\right]\right\}, 
\end{aligned}
\ee
\be\label{I3b1}
\begin{aligned}
\mathrm{I_3}_{(\ell,\ell)}^{(\infty,\infty)}&=\frac
{A_f\pi}{2}B_{g}^2\xi\exp(-2\ell/\xi)\int_0^\infty\!dy\frac{
\exp(-y/\xi)} {\left (y^2+1\right) ^2},
\end{aligned}
\ee
and
\be\label{I3c1}
\begin{aligned}
\mathrm{I_3}_{(d_w,\ell)}^{(\ell,\infty)}&=\frac
{A_f\pi}{2}\left\{A_{l}B_{g}\left[\ell\left(\int_0^\infty\!\!\!dy\frac{
\exp(-y/\xi)}{\left(y^2+1\right)^2}-\int_0^{\ell-d_w}\!\!\!dy\frac{
\exp(-y/\xi)}{\left(y^2+1\right)^2}\right)+\int_0^{\ell-d_w}\!\!\!dy\frac{y
\exp(-y/\xi)}{\left(y^2+1\right)^2}\right.\right.\\
&\left.-d_w\int_{l-d_w}^\infty\!dy\frac{
\exp(-y/\xi)}{\left(y^2+1\right)^2}\right] 
+B_{l}B_{g}\frac{\xi}{2}\left[\exp(-2d_w/\xi)\int_{\ell-d_w}^\infty\!dy\frac{
\exp(-y/\xi)}{\left(y^2+1\right)^2}\right.\\
&\left.\left.+\exp(-2\ell/\xi)\left(\int_0^{\ell-d_w}\!dy\frac{
\exp(y/\xi)}{\left(y^2+1\right)^2}-\int_0^\infty\!dy\frac
{\exp(-y/\xi)}{\left(y^2+1\right)^2}\right)\right]\right\}. 
\end{aligned}
\ee
We note the following relations: 
\be
\begin{aligned}
A_{l}^2&=A_1^2\exp(-2\ell/\xi)\\
A_{l}B_{l}&=A_1B_1\exp(-2\ell/\xi)+A_1B_2\exp(-\ell/\xi)\\
A_{l}B_{g}&=A_1B_1\exp(-2\ell/\xi)+A_1B_2\exp(-\ell/\xi)-A_1^2\\
B_{l}^2&=B_1^2\exp(-2\ell/\xi)+2B_1B_2\exp(-\ell/\xi)+B_2^2\\
B_{l}B_{g}
&=B_1^2\exp(-2\ell/\xi)+2B_1B_2\exp(-\ell/\xi)-A_1B_2\exp(\ell/\xi)+B_2^2-A_1B_1\\
B_{g}
^2&=B_1^2\exp(-2\ell/\xi)+2B_1B_2\exp(-\ell/\xi)-2A_1B_2\exp(\ell/\xi)\\
&+A_1^2\exp(2\ell/\xi)+B_2^2-2A_1 B_1.
\end{aligned}
\ee
Accordingly we obtain
\be
\begin{aligned}
 \frac12\left(\mathrm{I_3}_{(d_w,d_w)}^{(\ell,\ell)}\!+\!\mathrm{I_3}_{(d_w,\ell)}^{(\ell,\infty)}\!
+\!\mathrm { I_3} _
{(\ell,d_w)}^{(\infty,\ell)}\!+\!\mathrm{I_3}_{(\ell,\ell)}^{(\infty,\infty)}\right)&=\frac
{A_f\pi}{2}\left\{\frac{A_{l}^2\xi}{2}\left[\exp(2\ell/\xi)\int_{0}^{\ell-d_w}\!\!dy\frac{
\exp(-y/\xi)} {\left(
y^2+1\right)^2}\right.\right.\\
&\left.-\exp(2d_w/\xi)\int_{0}^{\ell-d_w}\!\!\!\!dy\frac{
\exp(y/\xi)} { \left(y^2+1\right)^2}\right]\\
&-A_1^2\left[(\ell\!-\!d_w)\!\!\int_{\ell-d_w}^\infty\!\!\!\!\!\!\!\!dy\frac{
\exp(-y/\xi)}{\left(y^2+1\right)^2}\!+\!\int_0^{\ell-d_w}\!\!\!\!\!\!\!dy\frac{y
\exp(-y/\xi)}{\left(y^2+1\right)^2}\right]\\
&+A_{l}B_{l}\left[
(\ell\!-\!d_w)\!\!\int_{d_w-\ell}^\infty\!\!\!\!\!\!\!\!dy\frac {
\exp(-y/\xi)}{\left(y^2+1\right)^2}\!-\!\!\int_0^{\ell-d_w}\!\!\!\!\!\!\!dy\frac {y
\exp(y/\xi)}{\left(y^2+1\right)^2}\right]\\
&+\frac{B_{l}^2\xi}{2}\exp(-2d_w/\xi)\int_{0}^\infty\!dy\frac{
\exp(-y/\xi)} { \left(y^2+1\right)^2}\\
&+\left(-A_1B_1-A_1B_2\exp(\ell/\xi)\right)\frac{\xi}{2}\times\\
&\times\left[\exp\left(-2d_w/\xi\right)\int_{\ell-d_w}^\infty\!\!dy\frac{
\exp\left(-y/\xi\right)}{\left(y^2+1\right)^2}\right.\\
&+\left.\exp\left(-2\ell/\xi\right)\int_0^{\ell-d_w}\!\!\!dy\frac{
\exp(y/\xi)}{\left(y^2+1\right)^2}\right]\\
&+\frac{\xi}{2}\left(A_1^2\exp(2\ell/\xi)-A_1B_2\exp(\ell/\xi)-A_1B_1\right)\times\\
&\times\left.\exp(-2\ell/\xi)\int_0^\infty\!dy
\frac{\exp(-y/\xi)} {\left (y^2+1\right) ^2}\right\}. 
\end{aligned}
\ee
In order to determine the asymptotic behavior of the integrals in Eqs. (\ref{I3a1})-(\ref{I3c1}) we
repeatedly integrate by parts so that

\be
\begin{aligned}
\int_{0}^{\ell-d_w}\!dy\frac{\exp(y/\xi)}{\left(y^2+1\right)^2}&=\frac{\xi\exp((\ell-d_w)/\xi)}{
\left(({\ell-d_w})^2+1\right)^2}-\xi+\xi\int_0^{\ell-d_w}\frac{4y}{\left(y^2+1\right)^3}\exp(y/\xi),
\\
&=\frac{\xi\exp((\ell-d_w)/\xi)}{\ell^4}+\mathcal{O}(\ell^{-5}\exp(\ell/\xi)),\ \ \
\ell\gg1,
\end{aligned}
\ee
\be
\begin{aligned}
\int_{0}^{\ell-d_w}\!dy\frac{\exp(-y/\xi)}{\left(y^2+1\right)^2}=\int_{0}^\infty\!dy\frac{
\exp(-y/\xi) } {
\left(y^2+1\right)^2}-\int_{\ell-d_w}^\infty\!dy\frac{\exp(-y/\xi)}{
\left(y^2+1\right)^2}
\end{aligned}
\ee
with
\be
\begin{aligned}
\int_{\ell-d_w}^\infty\!dy\frac{\exp(-y/\xi)}{\left(y^2+1\right)^2}&=\frac{\xi\exp(-(\ell-d_w)/\xi)}
{\left((\ell-d_w)^2+1\right)^2}-\xi\int_\ell^\infty\frac{4y}{\left(y^2+1\right)^3}\exp(-y/\xi)\\
&=\frac{\xi\exp(-(\ell-d_w)/\xi)}{\ell^4}+\mathcal{O}(\ell^{-5}\exp(-\ell/\xi)),\ \ \
\ell\gg d_w,
\end{aligned}
\ee
and
\be
\begin{aligned}
\int_{d_w-\ell}^\infty\!dy\frac{\exp(-y/\xi)}{\left(y^2+1\right)^2}&=\frac{
\xi\exp\left((\ell-d_w)/\xi\right) } {
\left((\ell-d_w)^2+1\right)^2}-\xi\int_{d_w-\ell}^\infty\frac{4y}{\left(y^2+1\right)^3}
\exp(-y/\xi)\\
&=\frac{\xi\exp((\ell-d_w)/\xi)}{\ell^4}+\mathcal{O}(\ell^{-5}\exp(\ell/\xi)),\ \ \ \ell\gg d_w.
\end{aligned}
\ee
Finally, collecting the leading terms for $\ell\to\infty$ one obtains the asymptotic behavior
\be\label{I3ar}
\begin{aligned}
\frac12\left(\mathrm{I_3}_{(d_w,d_w)}^{(\ell,\ell)}+\mathrm{I_3}_{(d_w,\ell)}^{(\ell,\infty)}
+\mathrm { I_3} _
{(\ell,d_w)}^{(\infty,\ell)}+\mathrm{I_3}_{(\ell,\ell)}^{(\infty,\infty)}\right)&=\frac
{A_f\pi\xi}{2}\left(A_1^2\!+\!\frac{B_2^2\exp\left(-\frac{2d_w}{\xi}\right)}{2}\right)\int_ { 0 }
^\infty\!\!\!dy\frac{\exp\left(-\frac{y}{\xi}\right)} { \left(
y^2+1\right)^2}\\
&-\frac{A_f\pi A_1^2}{2}\left(\frac{1}{2}\!-\!\frac{1}{2\xi}\int_0^\infty\!dy\frac{\exp(-y/\xi) }
{y^2+1}\right)\!+\!\mathcal{O}\left(\frac{1}{\ell^4}\right).
\end{aligned}
\ee

Inserting the results for these integrals (see Eqs.
(\ref{I11r}), (\ref{I12r}), (\ref{I21r}), (\ref{I22r}), (\ref{I2ar}), (\ref{I2br}), and
(\ref{I3ar})) into Eq. (\ref{func_approx-lg}), one obtains the effective interface
potential $\omega(\ell)=\Omega_{s,lr}(\ell)-\Omega_{s,lr}(\infty)$ given by Eq.
(\ref{omega2}); the index $lr$ refers to \textit{l}ong-\textit{r}anged interactions (Sec. \ref{lr}).
\end{widetext}



\end{document}